\documentclass[letterpaper,reprint,amsmath,amssymb,aps,prd,floatfix,showpacs]{revtex4-1}
\usepackage{graphicx}
\usepackage{dcolumn}
\usepackage{bm}
\usepackage{amssymb}
\usepackage{multirow}
\usepackage{epsfig}
\usepackage{epstopdf}
\usepackage{amsmath,booktabs,array}
\usepackage{tabularx}
\usepackage{slashbox}
\usepackage[hyperindex,breaklinks]{hyperref}
\usepackage{atlasphysics}
\usepackage{preprintcover}
\DeclareGraphicsExtensions{.eps}

\newcommand{\aidatitle}{{Simultaneous measurements of the \ttbar{}, \WWpm{}, and \Ztau{} production cross-sections in $pp$ collisions at $\rts = 7\TeV$ with the ATLAS detector}}

\newcommand{\Nfitfid}{\ensuremath{N^{\mathrm{fid}}_{X}}}

\newcommand{\Nfittot}{\ensuremath{N^{\mathrm{tot}}_{X}}}

\newcommand{\ttbarXsecTot}{\ensuremath{181.2}}
\newcommand{\ttbarUncertStat}{\ensuremath{\pm 2.8}}

\newcommand{\ttbarUncertLum}{\ensuremath{\pm 3.3}}
\newcommand{\ttbarUncertBE}{\ensuremath{\pm 3.3}}

\newcommand{\ttbarXsecFid}{\ensuremath{2730}}
\newcommand{\ttbarUncertStatFid}{\ensuremath{\pm 40}}

\newcommand{\ttbarUncertLumFid}{\ensuremath{\pm 50}}
\newcommand{\ttbarUncertBEFid}{\ensuremath{\pm 50}}

\newcommand{\ttbarXsecFidnotau}{\ensuremath{2374}}
\newcommand{\ttbarUncertStatFidnotau}{\ensuremath{\pm 37}}

\newcommand{\ttbarUncertLumFidnotau}{\ensuremath{\pm 43}}
\newcommand{\ttbarUncertBEFidnotau}{\ensuremath{\pm 43}}

\newcommand{\WWXsecTot}{\ensuremath{53.3}}
\newcommand{\WWUncertStat}{\ensuremath{\pm 2.7 }}

\newcommand{\WWUncertLum}{\ensuremath{\pm 1.0}}
\newcommand{\WWUncertBE}{\ensuremath{\pm 0.5}}

\newcommand{\WWXsecFid}{\ensuremath{638}}
\newcommand{\WWUncertStatFid}{\ensuremath{\pm 32 }}

\newcommand{\WWUncertLumFid}{\ensuremath{\pm 11}}
\newcommand{\WWUncertBEFid}{\ensuremath{\pm 6}}

\newcommand{\WWXsecFidnotau}{\ensuremath{563}}
\newcommand{\WWUncertStatFidnotau}{\ensuremath{\pm 28 }}

\newcommand{\WWUncertLumFidnotau}{\ensuremath{\pm 10}}
\newcommand{\WWUncertBEFidnotau}{\ensuremath{\pm 6}}

\newcommand{\ZtauXsecTot}{\ensuremath{1174}}
\newcommand{\ZtauUncertStat}{\ensuremath{\pm 24}}

\newcommand{\ZtauUncertLum}{\ensuremath{\pm 21}}
\newcommand{\ZtauUncertBE}{\ensuremath{\pm 9}}

\newcommand{\ZtauXsecFid}{\ensuremath{1690}}
\newcommand{\ZtauUncertStatFid}{\ensuremath{\pm 35}}

\newcommand{\ZtauUncertLumFid}{\ensuremath{\pm 30}}
\newcommand{\ZtauUncertBEFid}{\ensuremath{\pm 14}}

\newcommand{\lum}{\ensuremath{\cal{L}}}

\newcommand{\lumi}{4.6~\ifb}
\newcommand{\njets}{\ensuremath{N_{\mathrm{jets}}}}

\newcommand{\topmass}{\ensuremath{172.5\GeV}}

\renewcommand{\Ztau}{\ensuremath{Z/\gamma^{*}\rightarrow\tau\tau}}
\newcommand{\Wt}{\ensuremath{Wt}}
\newcommand{\WW}{\ensuremath{WW}}
\newcommand{\WWpm}{\ensuremath{W^+W^-}}

\newcommand{\ZZ}{\ensuremath{ZZ}}
\newcommand{\WZ}{\ensuremath{WZ}}
\newcommand{\emu}{\ensuremath{e\mu}}

\renewcommand{\mumu}{\ensuremath{\mu\mu}}

\newcommand{\A}{\ensuremath{\mathcal{A}}}

\newcommand{\C}{\ensuremath{\mathcal{C}}}

\newcommand{\metnjets}{\met{}--\njets{}}
\newcommand{\alpgen}{{\sc alpgen}}

\newcommand{\jimmy}{{\sc jimmy}}
\newcommand{\mcnlo}{{\sc mc@nlo}}
\newcommand{\powheg}{{\sc powheg}}

\newcommand{\pythia}{{\sc pythia}}

\newcommand{\herwig}{{\sc herwig}}
\newcommand{\sherpa}{{\sc sherpa}}
\newcommand{\geant}{{\sc geant4}}
\newcommand{\tauola}{{\sc tauola}}
\newcommand{\photos}{{\sc photos}}
\newcommand{\mcfm}{{\sc MCFM}}
\newcommand{\gftoww}{{\sc gg2WW}}

\newcommand{\toppp}{{\sc top++}}

\newcommand{\ie}{{i.e.}}

\PreprintCoverPaperTitle{\aidatitle}
\PreprintIdNumber{CERN-PH-EP-2014-119}
\PreprintCoverAbstract{
Simultaneous measurements of the $t\bar{t}$, $W^+W^-$, and $Z/\gamma^{*}\rightarrow\tau\tau$ production cross-sections using an integrated
luminosity of $4.6\,\mathrm{fb}^{-1}$ of $pp$ collisions at $\sqrt{s} = 7\,\mathrm{TeV}$ collected by the ATLAS detector at the LHC are presented. 
Events are selected with two high transverse momentum leptons consisting of an oppositely charged electron and muon pair.
The three processes are separated using the distributions of the missing transverse momentum of events with zero and greater than zero jet multiplicities.
Measurements of the fiducial cross-section are presented along with results that quantify for the first time the underlying correlations in the predicted and measured cross-sections due to proton parton distribution functions.
These results indicate that the correlated NLO predictions for $t\bar{t}$ and $Z/\gamma^{*}\rightarrow\tau\tau$ underestimate the data, while those at NNLO generally describe the data well.
The full cross-sections are measured to be $\sigma(t\bar{t})  = 181.2 \pm 2.8^{+9.7}_{-9.5} \pm 3.3 \pm 3.3\,\mathrm{pb}$, $\sigma(W^+W^-)  = 53.3 \pm 2.7^{+7.3}_{-8.0} \pm 1.0 \pm 0.5\,\mathrm{pb}$, and $\sigma(Z/\gamma^{*}\rightarrow\tau\tau)  = 1174 \pm 24^{+72}_{-87} \pm 21 \pm 9\,\mathrm{pb}$, where the cited uncertainties are due to statistics, systematic effects, luminosity and the LHC beam energy measurement, respectively. The $W^+W^-$ measurement includes the small contribution from Higgs boson decays, $H\rightarrow W^+W^-$.

}
\PreprintJournalName{PRD}

\begin{document}

\setlength{\tabcolsep}{8pt}

\title{\aidatitle}
\author{The ATLAS Collaboration}
\date{\today}

\begin{abstract}
Simultaneous measurements of the $t\bar{t}$, $W^+W^-$, and $Z/\gamma^{*}\rightarrow\tau\tau$ production cross-sections using an integrated
luminosity of $4.6\,\mathrm{fb}^{-1}$ of $pp$ collisions at $\sqrt{s} = 7\,\mathrm{TeV}$ collected by the ATLAS detector at the LHC are presented. 
Events are selected with two high transverse momentum leptons consisting of an oppositely charged electron and muon pair.
The three processes are separated using the distributions of the missing transverse momentum of events with zero and greater than zero jet multiplicities.
Measurements of the fiducial cross-section are presented along with results that quantify for the first time the underlying correlations in the predicted and measured cross-sections due to proton parton distribution functions.
These results indicate that the correlated NLO predictions for $t\bar{t}$ and $Z/\gamma^{*}\rightarrow\tau\tau$ underestimate the data, while those at NNLO generally describe the data well.
The full cross-sections are measured to be $\sigma(t\bar{t})  = 181.2 \pm 2.8^{+9.7}_{-9.5} \pm 3.3 \pm 3.3\,\mathrm{pb}$, $\sigma(W^+W^-)  = 53.3 \pm 2.7^{+7.3}_{-8.0} \pm 1.0 \pm 0.5\,\mathrm{pb}$, and $\sigma(Z/\gamma^{*}\rightarrow\tau\tau)  = 1174 \pm 24^{+72}_{-87} \pm 21 \pm 9\,\mathrm{pb}$, where the cited uncertainties are due to statistics, systematic effects, luminosity and the LHC beam energy measurement, respectively. The $W^+W^-$ measurement includes the small contribution from Higgs boson decays, $H\rightarrow W^+W^-$.

\end{abstract}
\pacs{13.85.Lg, 14.65.Ha, 14.70.Hp, 14.70.Fm}

\maketitle

\section{Introduction}
\label{Sec:Intro}
Proton collisions at the LHC have large cross-sections for the production of top quark pairs, $W$ boson pairs, and $Z$ bosons.
The cross-section for each of these processes is predicted to a high precision within the standard model (SM) of particle physics.
In this article, a global test of these SM predictions is presented through the study of a common final state including an oppositely charged electron and muon pair ($e\mu$ events).
Specifically, a simultaneous measurement of the cross-sections of the pair production of top quarks (\ttbar), $W$ bosons ($W^+W^-$, written as \WW), and tau-leptons via 
the Drell--Yan mechanism (\Ztau) is performed.
These processes are considered in a two-dimensional parameter space spanned by the missing transverse momentum, \met{}, and jet multiplicity, \njets{}, where they are naturally well separated, allowing the simultaneous extraction of their cross-sections.
Events from \ttbar{} production tend to have large \met{} and large \njets{}, whereas \WW{} events tend to have large \met{} and small \njets{}, and \Ztau{} events are characterized by small \met{} and even smaller \njets{}.

This analysis of $e\mu$ events allows a broader test of the SM than 
that given by dedicated cross-section measurements, and provides
a first simultaneous measurement of the production cross-sections for the processes of interest at the LHC.
This simultaneous measurement unifies the definitions of fiducial region, physics object and event selections, and estimation of uncertainties for each signal measurement.
In particular these measurements offer a new window on the effects of the parton distribution functions (PDF) through consideration of the correlations between pairs of production cross-sections, induced by the use of common PDF predictions.
An improved understanding of these processes can improve the theoretical calculations and methods used in their study, and thereby more precisely constrain background predictions for future new physics searches at the LHC.

The measurement technique used here was first used by the CDF
experiment at the Tevatron~\cite{CDFpaper} using the $p\bar{p}$ collision data at a
center-of-mass energy, $\sqrt{s}$, of 1.96 TeV. In this paper the results are obtained from
$\sqrt{s} = 7\,{\rm TeV}$ $pp$ collision data collected by the ATLAS
detector~\cite{atlas} at the LHC corresponding to an integrated luminosity of
 \lumi{}~\cite{atlaslumi7TeV2}. Furthermore the measurement of \WW{} includes the small contribution from Higgs boson decays, $H\rightarrow W^+W^-$. 
Previous dedicated measurements of these cross-sections in the dilepton channel were performed by ATLAS using data samples of \lumi{}~for
\ttbar{}~\cite{Aad:2014kva} and $WW$~\cite{CERN-PH-EP-2012-242}, and $36~\ipb$~for \Ztau{}~\cite{Aad:2011kt}. Other dedicated measurements in the dilepton channel were also performed by the CMS collaboration, namely for \ttbar{} using $2.3~\ifb$~\cite{Chatrchyan:2012bra}, for \WW{} using $4.9~\ifb$~\cite{Chatrchyan:2013yaa}, and for \Ztau{}~\cite{Chatrchyan:2011nv}  using $36~\ipb$. 

This paper is organized as follows. 
Section~\ref{Sec:ATLASDet} provides an overview of the ATLAS detector.
Section~\ref{Sec:MC} describes the data sample and summarizes the Monte Carlo simulation used for the key SM processes relevant to this study, while Sec.~\ref{Sec:ObjEvtSelec} details the reconstruction of the final-state objects, the \emu{} event selection, as well as the full definition of the \metnjets{} parameter space.
Section~\ref{Sec:DDB} covers the data-driven estimation of backgrounds from misidentified and non-prompt leptons.
The template fitting method used to extract the results is discussed in Sec.~\ref{Sec:FitMeth} along with the treatment and evaluation of systematic uncertainties.
Results obtained for the cross-sections of the three processes of interest are presented and compared to predictions and other measurements in Sec.~\ref{Sec:Xsec}, and conclusions are presented in Sec.~\ref{Sec:conclusion}.
\section{The ATLAS Detector}
\label{Sec:ATLASDet}

ATLAS~\cite{atlas} is a multi-purpose particle physics detector with forward-backward symmetric cylindrical geometry.
The inner detector (ID) system is immersed in a 2 T axial magnetic field and provides tracking information for charged particles in the pseudorapidity range $|\eta|< 2.5$~\footnote{The ATLAS reference system is a right-handed Cartesian co-ordinate system, with the nominal collision point at the origin. The anti-clockwise beam direction defines the positive $z$-axis, while the positive $x$-axis is defined as pointing from the collision point to the center of the LHC ring and the positive $y$-axis points upwards. The azimuthal angle, $\phi$, is measured around the beam axis, and the polar angle, $\theta$, is the angle measured with respect to the $z$-axis. The pseudorapidity is given by $\eta = -\ln \tan(\theta/2)$. Transverse momentum and energy are defined as $p_{\text{T}} = p \sin\theta$ and $E_{\text{T}} = E \sin\theta$, respectively.}. 
It consists of a silicon pixel detector, a silicon microstrip detector, and a transition radiation tracker (TRT). 

The calorimeter system covers the range $|\eta|< 4.9$. 
The highly segmented electromagnetic calorimeter consists of lead absorbers with liquid-argon (LAr) as active material and covers the range $|\eta|< 3.2$. 
In the region $|\eta|< 1.8$, a pre-sampler detector using a thin layer of LAr is used to correct for the energy lost by electrons and photons upstream of the calorimeter. 
The hadronic tile calorimeter is a steel/scintillator-tile detector and is situated directly outside of the electromagnetic calorimeter. The barrel
section of this sampling calorimeter provides a coverage of $|\eta|< 1.7$. The endcap hadronic calorimeters have LAr as the active material
and copper absorbers covering the range  $1.5 < |\eta| < 3.2$. They cover the region between
the barrel and the forward calorimeter with a small overlap with each of them. The
forward calorimeter uses LAr as active material and copper and tungsten as absorber materials.
It extends the calorimeter coverage out to  $|\eta| = 4.9$.

The muon spectrometer (MS) measures the deflection of muons in the magnetic field produced by the large superconducting air-core toroid magnets. 
It covers the range $|\eta|< 2.7$ and is instrumented with separate trigger and high-precision tracking chambers. 
A precision measurement of the track coordinates in the bending direction of the toroidal magnetic field is provided by drift tubes in the range $|\eta|< 2.7$.
Within the region $2.0 < |\eta|< 2.7$, cathode strip chambers with higher granularity are used in the inner-most tracking layer. 
The muon trigger system, which covers the range $|\eta|< 2.4$, consists of resistive plate chambers in the barrel ($|\eta|< 1.05$) and thin gap chambers in the endcap regions ($1.05 < |\eta|< 2.4$).

A three-level trigger system is used to select events for offline analysis. 
The level-1 trigger is implemented in hardware and uses a subset of the detector information to reduce the event rate to its design value of at most 75~kHz. 
This is followed by two software-based trigger levels, level-2 and the event filter, which together reduce the event rate
to an average of 400 Hz during the 2011 data-taking period.
\section{Data and Monte Carlo Samples}
\label{Sec:MC}

The data sample used in this measurement consists of proton--proton collision events at a center-of-mass energy $\sqrt{s}= 7$~\TeV{} recorded by ATLAS in 2011.
Only data collected during stable beam conditions and with the relevant ATLAS sub-systems operational are used.
In particular, the inner detector, the electromagnetic and hadronic calorimeters, and the muon spectrometer must deliver data of high quality to ensure that electrons, muons, jets, and missing transverse momentum are measured accurately. The data selected for this study were collected 
using single-lepton triggers ($e$ or $\mu$).
In the case of the electron trigger, a threshold is applied to the transverse energy ($\et$) of the electron while for the muon trigger a threshold is applied to the transverse momentum ($\pt$)  of the muon. Due to the increases in luminosity achieved by the LHC during the 2011 run, the value of the electron $\et$ threshold applied changed during the course of the year. Thresholds employed by the electron trigger were
either 20 GeV or 22 GeV while the muon trigger threshold remained constant at 18 GeV. The data collected correspond to 
an integrated luminosity of \lumi{}, after applying data quality requirements, with an uncertainty of 1.8\%~\cite{atlaslumi7TeV2}. 

Monte Carlo simulated events are generated at $\sqrt{s}=7$~TeV and processed through a detector simulation~\cite{atlassim} based on \geant{}~\cite{Agostinelli:2002hh}. In these samples, all particle masses are taken from 2010 values published by the Particle Data Group~\cite{PDG2010} with the exception of the top quark mass, which is taken to be \topmass{} and the Higgs boson mass which is set to $125\,\GeV$.
The simulation includes modeling of additional $pp$ interactions in the same and neighboring bunch crossings, referred to as pile-up.
These events are subsequently reweighted such that the distribution of the number of interactions per bunch crossing in simulation matches that of data.
Corrections to the selection efficiency of electrons and muons are applied to simulated events, and the detector simulation is tuned to reproduce the energy and momentum measurements and resolution observed in data.

Unless otherwise specified, common attributes between the Monte Carlo samples are the generation of the underlying event (UE), which is performed by \pythia{} v.~6.425~\cite{Pythia64} or \jimmy{} v.~4.31~\cite{MultipartonIntPhoto}
(included as part of the \herwig{} v.~6.520~\cite{Herwig6.5} software package), and the choice of PDFs, which is the NLO
CT10 set~\cite{CT10PDF}. An exception is the \alpgen{}~\cite{Alpgen2002} generator configurations which use the leading-order set CTEQ6L1~\cite{CTEQ6L1}.

The cross-sections for the different processes obtained from a range of event generators are always normalized to the best available theoretical calculations, as discussed below.

\subsection{\ttbar{}~production}
Simulation of \ttbar{} production is performed using the next-to-leading-order (NLO) generator \mcnlo{} v4.01~\cite{MC@NLO2002} interfaced to \herwig{} and \jimmy{}.
The \ttbar{} cross-section has been calculated at next-to-next-to-leading-order (NNLO) in QCD, including resummations of next-to-next-to-leading logarithmic (NNLL) soft gluon terms with \toppp 2.0~\cite{Cacciari:2011hy,Baernreuther:2012ws,Czakon:2012zr,Czakon:2012pz,Czakon:2013goa,Top++2011}. The resulting cross-section is calculated to be $\sigma_{\ttbar}=177^{+10}_{-11}$~pb for a top quark mass of $172.5$ \GeV~\cite{CDFandD0:2008aa}. 
The uncertainty due to the choice of PDF and $\alpha_{s}$ is calculated using the PDF4LHC prescription~\cite{PDF4LHC}  that includes 
the MSTW2008 68\% CL NNLO~\cite{MSTW2008,Martin:2009bu}, CT10 NNLO~\cite{CT10PDF,Gao:2013xoa} and NNPDF2.3 5f FFN~\cite{Ball:2012cx} PDF sets. This is
added in quadrature to the scale uncertainty. 

Additional samples are provided using \powheg{}~\cite{Powheg2010} version powheg-hvq4 interfaced to the \pythia{} and \herwig{} parton shower generators, to compare parton shower (PS) and fragmentation models, and to assign a generator modeling uncertainty.

To estimate uncertainties due to modeling of QCD initial- (ISR) and final-state radiation (FSR) in the \ttbar{} system (discussed in Sec.~\ref{Sec:FitMeth}), \alpgen{} interfaced to the \pythia{} PS generator is used.
The uncertainty is evaluated using two different generator tunes with increased or reduced rates of QCD radiation.

\subsection{\WW{}~production}
The simulation of \WW{} signal production is based on samples of $q\bar{q} \to WW$, $gg \to WW$ and $gg \to H \to WW$ events, which are generated with \mcnlo{}, \gftoww{}~\cite{Binoth:2006mf}, and \powheg{} respectively. The Higgs resonance sample is interfaced to \pythia{} and the non-resonant samples are interfaced to \herwig{}. A combined \WW{} sample is formed from cross-section weighted contributions, where cross-sections of $44.7^{+2.1}_{-1.9}$~pb, $1.3^{+0.8}_{-0.5}$~pb and $3.3 \pm 0.3$~pb are assumed for $q\bar{q} \to WW$, $gg \to WW$ and $gg \to H \to WW$, respectively~\cite{Campbell:2009kg,Heinemeyer:2013tqa}. 

Alternative \WW{} samples are produced with the \powheg{} generator interfaced to \pythia{} and \herwig{} PS generators for comparison of PS and fragmentation models and to assess a generator modeling uncertainty. \alpgen~samples are used to
estimate uncertainties due to modeling of additional QCD radiation.

\subsection{Drell--Yan lepton pair production}
The only Drell--Yan process whose final states include a prompt $e$ and $\mu$ is the production of a pair of tau-leptons. For \Ztau{}, 
the \sherpa{} v.~1.4.0~\cite{SHERPA2009} generator is used.
{\sc Sherpa} handles the full generation of the event, including a fixed-order matrix element calculation, parton showering, hadronization, and underlying event.
The cross-section for inclusive $Z/\gamma^*$ production is calculated at NNLO in {\sc FEWZ}~\cite{FEWZ2011} with MSTW2008 NNLO PDFs to be ${\sigma_{\mathrm{NNLO}}^{\Ztau} = 1070 \pm 54}$~pb.
This calculation is performed for $m_{\tau \tau} > 40\GeV$, and includes contributions from $\gamma^{*} \to \tau \tau$.

\subsection{Single top quark production}
The associated production of a single top quark and a W boson, referred to as the $Wt$ channel, is simulated with \mcnlo{} interfaced to \herwig{} and \jimmy{}. Single top production
through the $s$ and $t$ channels is not considered here, since only the $Wt$ channel is a source of prompt $e\mu$ pairs. These
are considered a background in the analysis.
During event generation a diagram removal scheme is implemented \cite{MC@NLOWt1,MC@NLOWt2} to remove overlaps between the single top and \ttbar{} final states. 
The cross-section for the $Wt$ channel calculated at approximate NNLO is ${\sigma_{\mathrm{theory}}^{Wt} = 15.7 \pm 1.1}$~pb~\cite{Kidonakis2010Wt}.

\subsection{$WZ$ and $ZZ$ production}
In the analysis, prompt $e\mu$ events originating from diboson samples, such as $WZ$ and $ZZ$, are considered part of the background. These are generated with \alpgen{} interfaced to \herwig{} and \jimmy{}. The NLO cross-sections for these processes are calculated with \mcfm{} v5.8~\cite{MCFM} with MSTW2008 NLO PDFs~\cite{MSTW2008}, and found to be ${\sigma_{\mathrm{NLO}}^{WZ} = 17.8 \pm 1.3}$~pb and ${\sigma_{\mathrm{NLO}}^{ZZ} = 5.9 \pm 0.3}$~pb for $m_Z > 60$~GeV.

\section{Object and Event Selection}
\label{Sec:ObjEvtSelec}

The high precision tracking of the ATLAS ID provides efficient reconstruction of multiple inelastic $pp$ collisions that take place 
in a single bunch crossing. The primary vertex is selected as the one with the largest sum of squared transverse momenta of
associated ID tracks. Contamination due to poorly reconstructed vertices is reduced by requiring that the primary vertex has at least five associated tracks with $\pt > 0.4 \GeV$.

Electron candidates are formed by an electromagnetic energy cluster with an associated track in the ID.
They must fulfill $|\eta| < 2.47$ with an exception of $1.37 < |\eta| < 1.52$ to exclude the transition region between the barrel and endcaps of the calorimeter. The candidates are required to have a transverse 
energy of $\et > 25 \GeV$ and meet the ``tight'' selection
criteria~\cite{Aad:2014fxa} optimized for the 2011 data-taking period. These criteria are based on 
the quality of the position and momentum association between the extrapolated track and the calorimeter energy cluster, the consistency of 
the longitudinal and lateral shower profiles with those expected for an incident electron, and the observed transition radiation in the TRT. To suppress background
 from photon conversions, the electron track is required to have a hit in the innermost layer of the tracking system.

Muon candidates are reconstructed by combining the information from pairs of stand-alone ID and MS tracks 
to form a single track~\cite{ATLAS:2010xza,ATLAS:2010qca}. The candidates are required to have $\pt > 20 \GeV$ and be located within 
the central region of the detector ($|\eta| < 2.5$).

The longitudinal impact parameter of each lepton with respect to the primary vertex is required to be less than 2 mm in order to suppress the non-prompt production of leptons.
To suppress the contribution from hadronic jets misidentified as leptons, electron and muon candidates are required to be isolated in both the ID and the calorimeter. 
Specifically, two measures of isolation are used: the sum of transverse energies of all calorimeter energy cells around the lepton but not associated with the lepton within a cone of size $\Delta R \equiv \sqrt{(\Delta \phi)^2 + (\Delta \eta)^2}=0.2$, denoted $\et^{\mathrm{cone20}}$, and the scalar sum of the transverse momenta of all tracks with $\pt > 1$ GeV that originate from the primary vertex and are within a cone of size $\Delta R =0.3$ around the lepton track, denoted $\pt^{\mathrm{cone30}}$. 
For electrons the maximum allowed values for $\et^{\mathrm{cone20}}$ and $\pt^{\mathrm{cone30}}$ are chosen as a function of the cluster $\eta$ so 
that the efficiency for the requirement measured in a $Z\to ee$ control sample is 90\% across the detector.
These values are also adjusted to account for pile-up conditions and energy leakage from the calorimeter.
The isolation requirement applied to the muons, $\et^{\mathrm{cone20}}< 4\,\mathrm{GeV}$ and $\pt^{\mathrm{cone30}}< 2.5\,\mathrm{GeV}$, has an overall efficiency of
96\% determined using a $Z\to \mu\mu$ control sample. 
The combination of cone sizes and efficiency working points was studied and optimized to find a requirement that reduces dependence on the pile-up conditions of the event. 

Jets are reconstructed using the anti-$k_{t}$ algorithm \cite{jetantikt} with a radius parameter of $R = 0.4$. 
The inputs to the jet algorithm are topological clusters of calorimeter cells. These topological clusters are seeded by calorimeter cells with energy $|E_{\mathrm{cell}}| > 4\sigma$, where $\sigma$ is the cell-by-cell RMS of the noise (electronics plus pile-up).
Neighboring cells are added if $|E_{\mathrm{cell}}| > 2\sigma$ and topological clusters are formed through an iterative procedure.
In a final step, all remaining neighboring cells are added to the topological cluster.
The baseline calibration for these topological clusters calculates their energy using the electromagnetic energy scale~\cite{Aad:2011he}.
This is established using test-beam measurements for electrons and muons in the electromagnetic and hadronic calorimeters~\cite{Aad:2014una,Aad:2010af}.

Effects due to non-compensation, energy losses in the dead material, shower leakage, as well as inefficiencies in energy clustering and jet reconstruction are also taken into account.
This is done by associating calorimeter jets with simulated jets in bins of $\eta$ and $E$, and is supplemented by an in-situ calibration.
This jet energy scale (JES) calibration is thoroughly discussed in Ref.~\cite{Aad:2014bia}.

To count a jet in the context of this analysis, it needs to fulfill the following kinematic requirements: $\pt > 30$ GeV and $|\eta| < 2.5$. 
A cut on the jet vertex fraction (JVF) is applied to minimize the number of jets originating from pile-up.
The JVF is defined as the ratio of the sum of the \pt{} of charged particle tracks that are associated with both the jet and the primary vertex, to the sum of the \pt{} of all tracks belonging to the jet. Its value must be greater than 75\%. 

To further remove non-prompt leptons that are likely to have originated from heavy-quark decays, leptons within a distance of
$\Delta R = 0.4$ from a reconstructed jet with $\pt > 25\,\mathrm{GeV}$ and $\mathrm{JVF} > 0.75$ are vetoed.

The second discriminating variable of the parameter space is the imbalance of the transverse momentum measured in each event due to the presence of neutrinos.
The reconstruction of the direction and magnitude (\met{}) of the missing transverse momentum vector is described in Ref.~\cite{Aad:2012re}.
It is calculated from the vector sum of the transverse momenta of all jets with $\pt > 20 \GeV$ and $|\eta| < 4.5$, the transverse momenta of electron and muon candidates, and finally from all calorimeter energy clusters not belonging to a reconstructed object.

Events are required to contain exactly one selected electron and one selected muon of opposite charge.
Events with an electron and muon of same-sign charge are used as a control sample for background studies.
Some properties of the electrons, muons, and jets belonging to events that satisfy the criteria described in this section are shown in 
Figs.~\ref{fig:pt_eta_lepton_histos} and~\ref{fig:mass_ht_histos}, where signal and background prompt processes are normalized to theory predictions
and the fake and non-prompt backgrounds are obtained as described in Sec.~\ref{Sec:DDB}. The data and simulation 
agree within the uncertainties associated with the theoretical predictions.

\begin{figure*}[htbp]
 \begin{center}
    \begin{tabular}{cc}
      \includegraphics[width=0.50\textwidth]{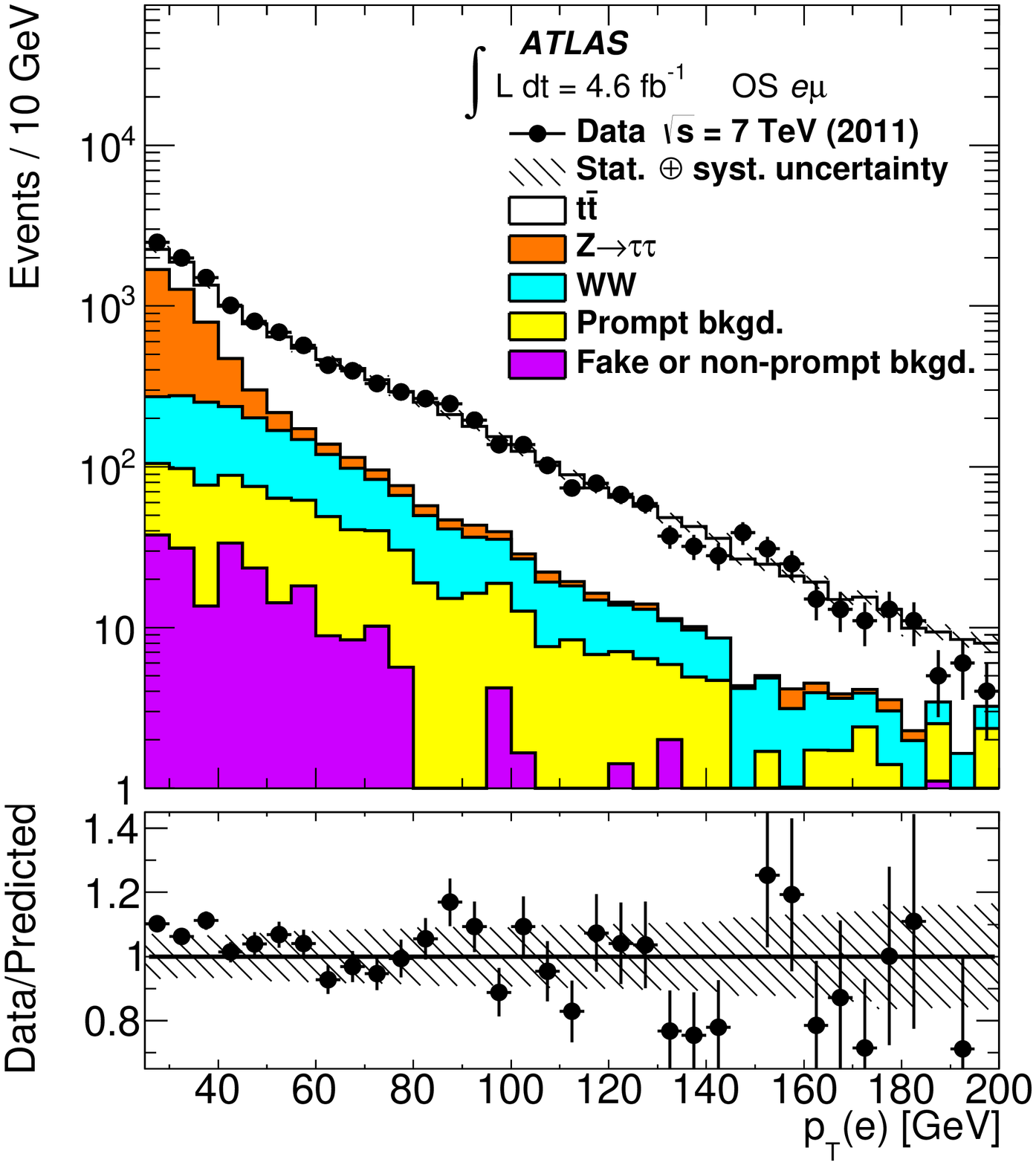} &
      \includegraphics[width=0.50\textwidth]{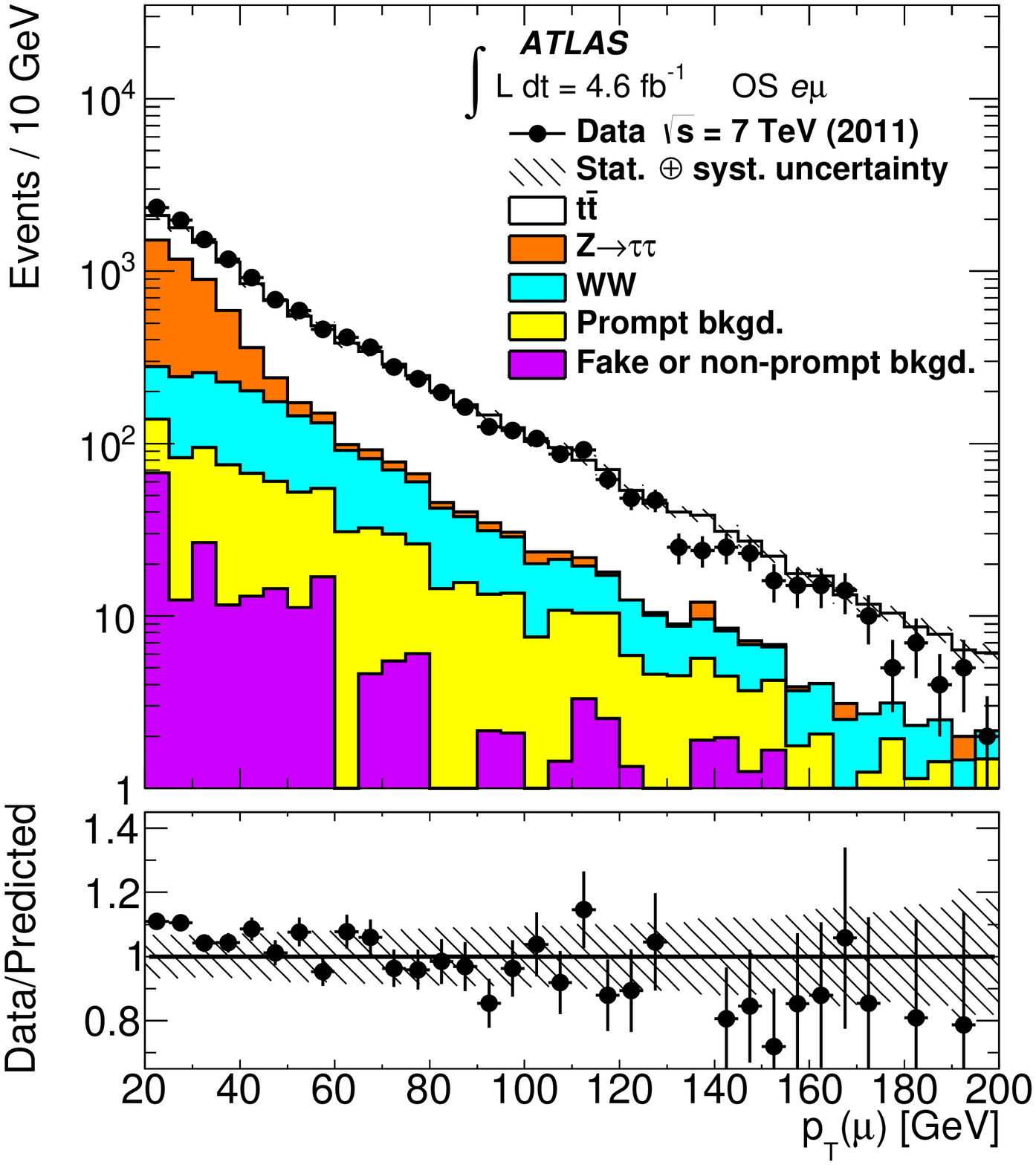}  \\
      (a) & (b) \\
      \includegraphics[width=0.50\textwidth]{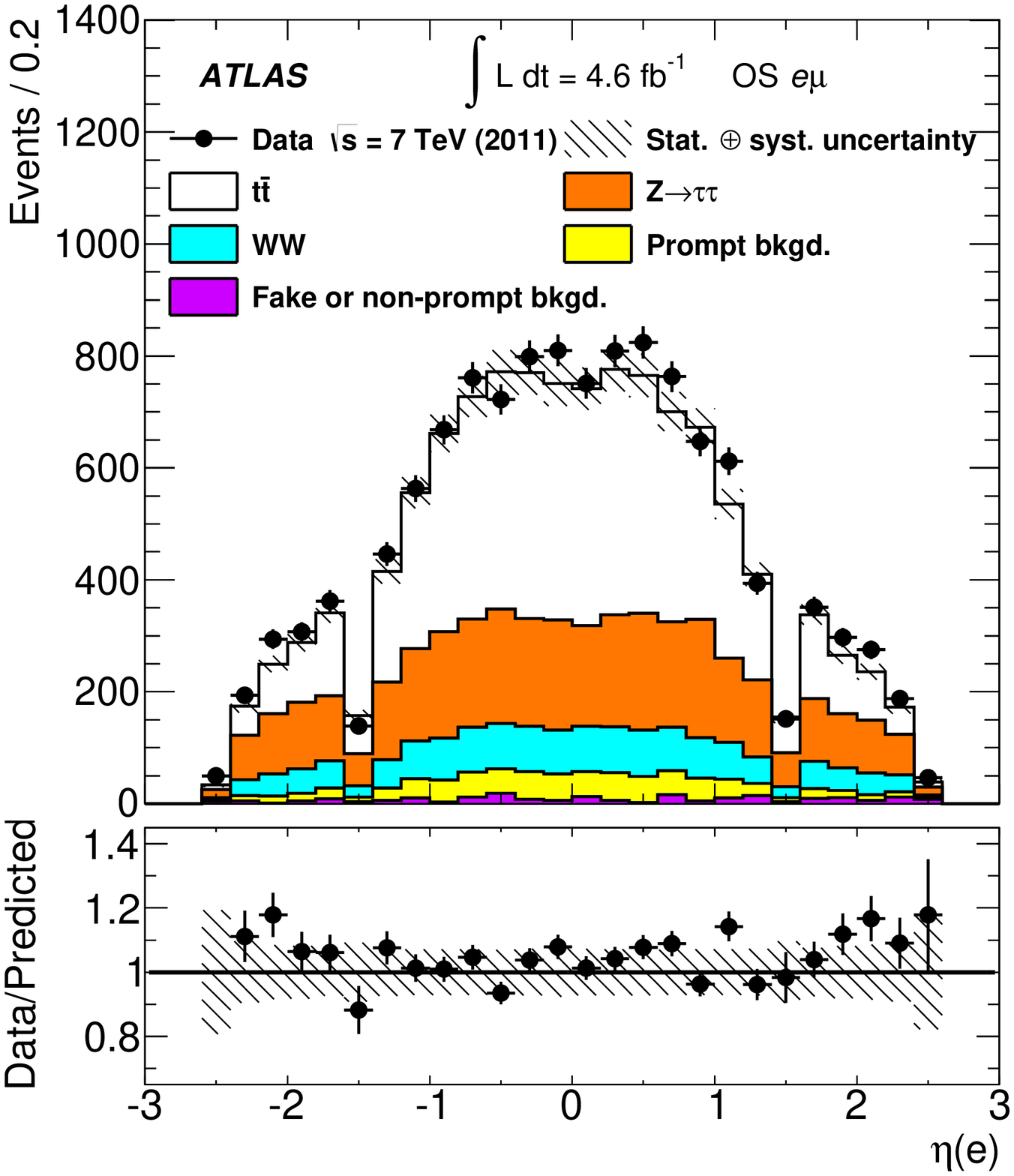} &
      \includegraphics[width=0.50\textwidth]{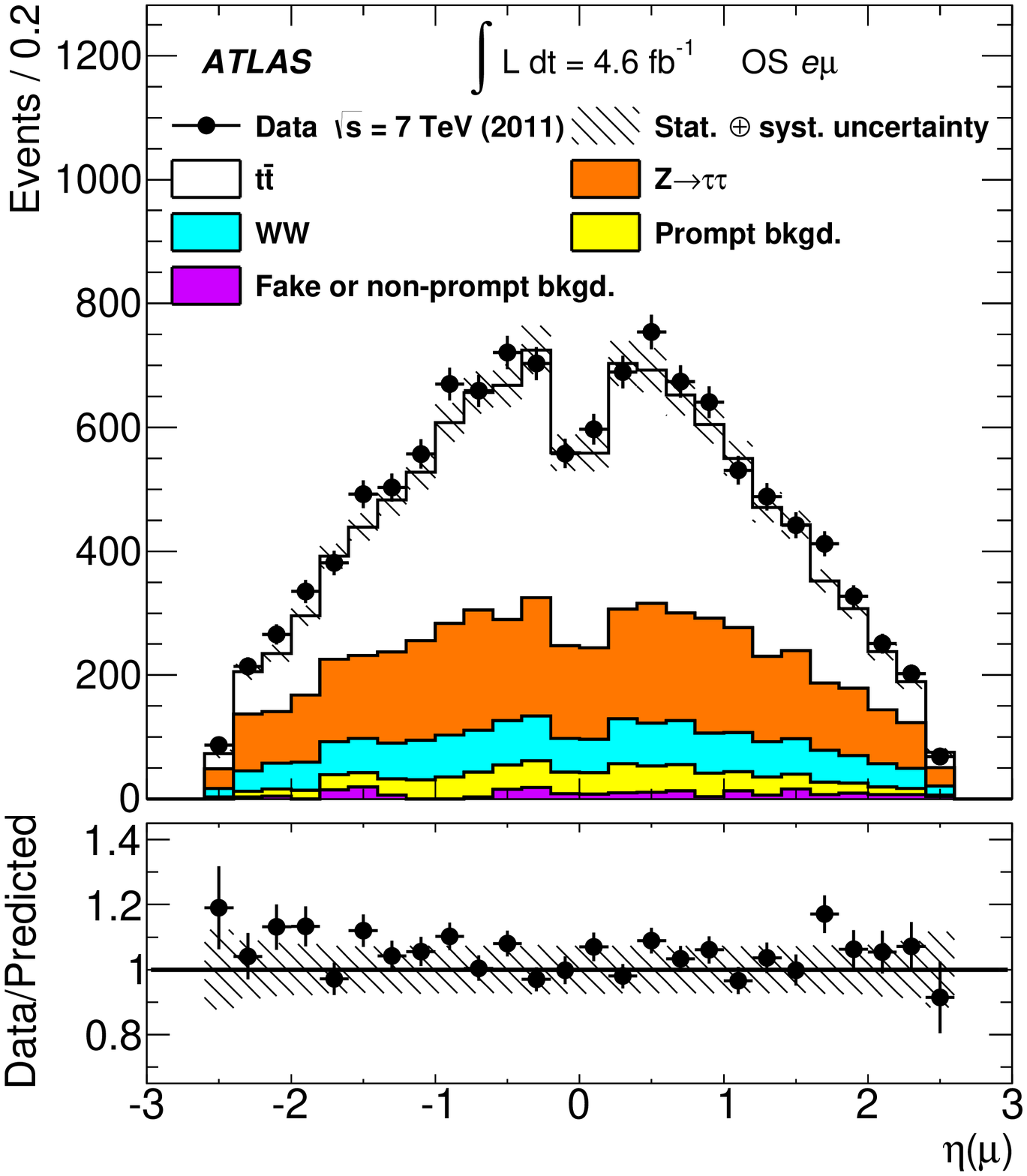}  \\
      (c) & (d) \\
    \end{tabular}
  \end{center}
\caption{Comparison between data and Monte Carlo samples (including the data-driven fake described in Sec.~\ref{Sec:DDB} 
and non-prompt backgrounds described in Sec.~\ref{Sec:MC}) normalized to their theoretical cross-sections
  for an integrated luminosity of \lumi{}: (a) electron and (b) muon candidate \pt{} distributions and, (c) and (d), their respective $\eta$ distributions
 for events producing one electron and one muon of opposite-sign (OS) charge.
    The electron and muon satisfy the signal region selection criteria presented in Sec.~\ref{Sec:ObjEvtSelec}. A bin by bin ratio between 
    the data and simulated events is shown at the bottom of each comparison.
    The hatched regions represent the combination of statistical and systematic uncertainties as listed in Table~\ref{t:sys_error} (except for shape uncertainties) and described in Sec.~\ref{Sec:FitMeth} together with the full theoretical cross-section uncertainties 
for the \ttbar{}, \WW{}, and \Ztau{} signal processes.
 \label{fig:pt_eta_lepton_histos}}
\end{figure*}

\begin{figure*}[htbp]
 \begin{center}
    \begin{tabular}{cc}
      \includegraphics[width=0.50\textwidth]{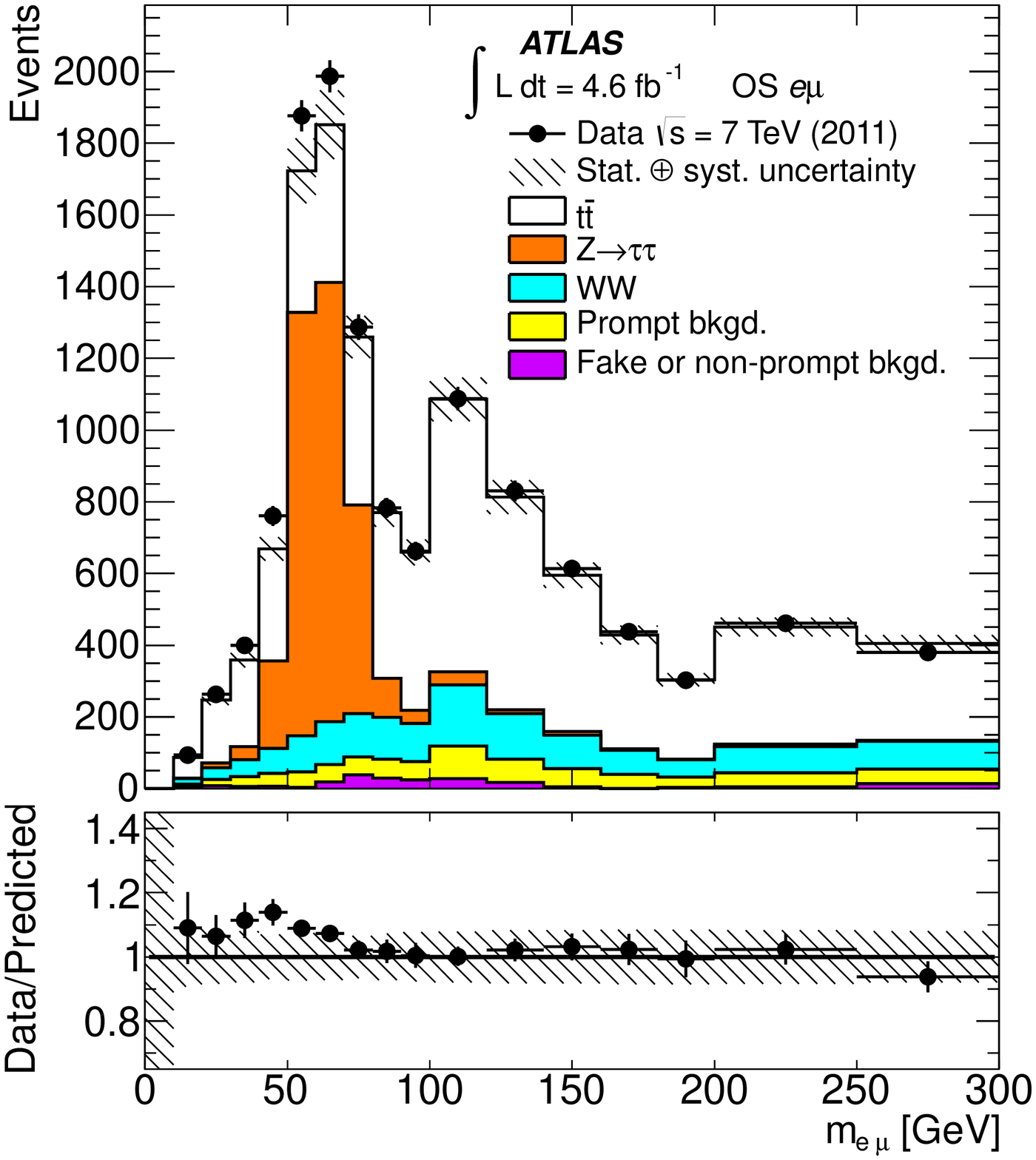} &
      \includegraphics[width=0.50\textwidth]{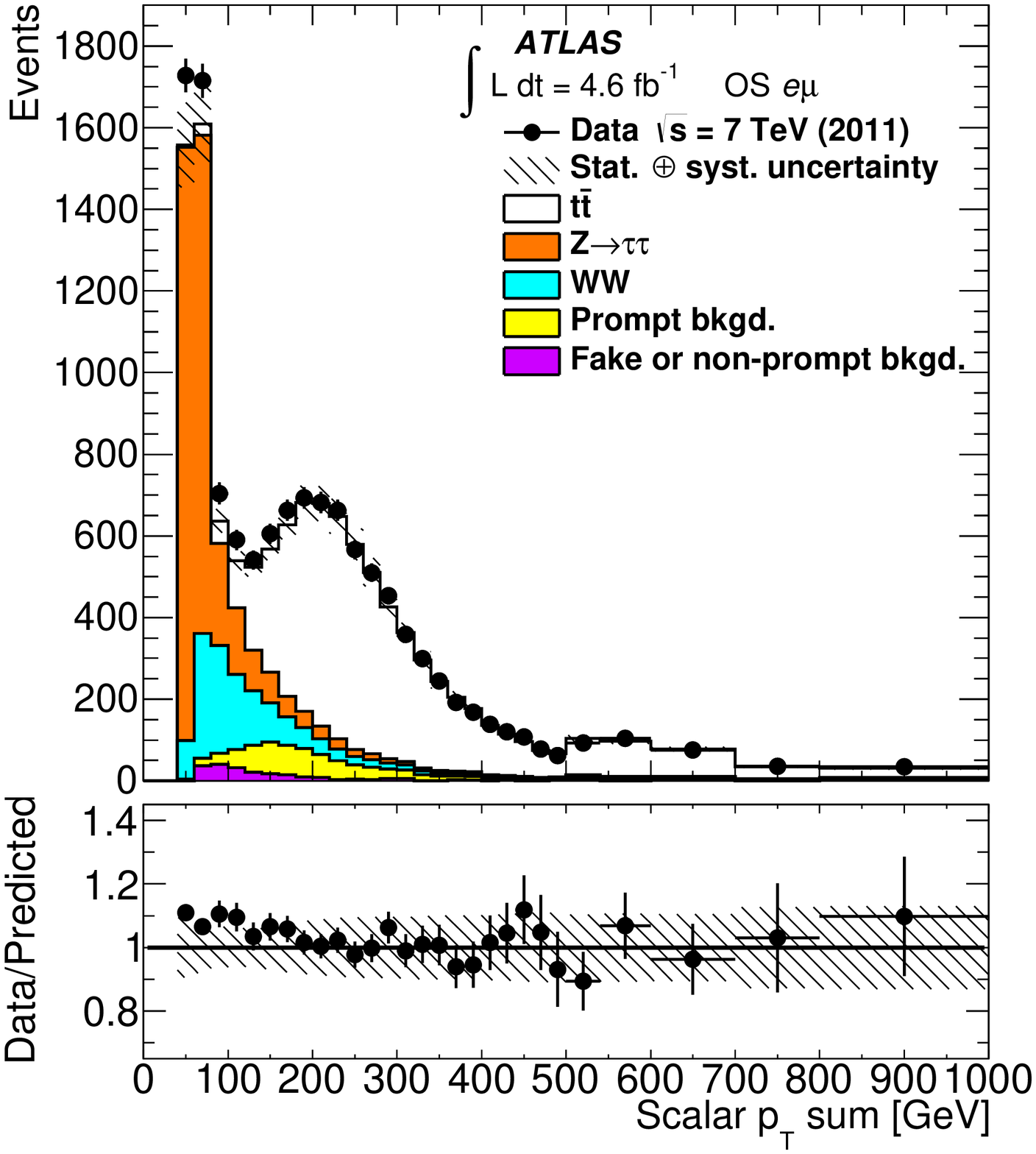}  \\
      (a) & (b) \\
    \end{tabular}
  \end{center}
\caption{Comparison between data and Monte Carlo samples (including the data-driven fake described in Sec.~\ref{Sec:DDB} 
and non-prompt backgrounds described in Sec.~\ref{Sec:MC}) normalized to their theoretical cross-sections
  for an integrated luminosity of \lumi{}: (a) invariant mass distribution of electron and muon pairs and (b) distribution of the scalar sum of the 
    transverse momenta of the
    selected electron, muon and jets. The electron and muon of opposite-sign (OS) charge satisfy the signal region 
    selection criteria presented in Sec.~\ref{Sec:ObjEvtSelec}.
 A bin by bin ratio between the data and simulated events is shown at the bottom of each comparison.
    The hatched regions represent the combination of statistical and systematic uncertainties as listed in Table~\ref{t:sys_error} (except for shape uncertainties) and described in Sec.~\ref{Sec:FitMeth} together with the full theoretical cross-section uncertainties for
    the \ttbar{}, \WW{}, and \Ztau{} signal processes.
 \label{fig:mass_ht_histos}}
\end{figure*}

\section{Background Estimated from Data}
\label{Sec:DDB}
Background contributions that include events where one or both of the leptons are fake or non-prompt are challenging to model with Monte Carlo simulation.
These events include a lepton from a heavy-flavor quark decay, a jet misidentified as a lepton, or an electron from a photon conversion.
These background contributions are difficult to estimate from simulation due to the potential mismodeling and limited knowledge of the relative composition of the background.
Additionally, the probability of accepting an event is small enough that the statistical uncertainty on the simulated sample becomes a serious concern. The analysis therefore relies on auxiliary measurements in data to obtain a robust estimate of background contributions shown in Table~\ref{t:fakes}, using 
the matrix method described in Ref.~\cite{Aad:2010ey}.

\subsection{Matrix method}
The matrix method utilizes data where the standard object selection requirements (referred to as {\it tight} criteria, see Sec.~\ref{Sec:ObjEvtSelec}) on either electron or muon or both candidates are relaxed (referred to as {\it loose} criteria).
The premise of this approach is that lepton candidates satisfying looser requirements have a higher chance of being fake or non-prompt than those satisfying {\it tight} requirements. This information combined with inputs of the probability that a real lepton or fake or non-prompt lepton meeting the {\it loose} criteria also satisfies the {\it tight} criteria, is used to arrive at a background estimate. For {\it loose} electrons, the isolation requirements are dropped, and electron identification criteria as defined in Ref.~\cite{Aad:2014fxa} are used, where the requirements on particle identification in the TRT and on the calorimeter energy to track momentum ratio $E/p$, are relaxed.  For {\it loose} muons, the isolation requirements are dropped.

For a given selected event, the matrix method, by solving a set of linear equations, implements a change of basis from observed data regions into event categories. The data regions comprise the signal region that is defined by a {\it tight} electron and a {\it tight} muon, denoted ``TT''; and control regions, containing events that produce a {\it tight} electron and a {\it loose} and not {\it tight} muon, denoted ``TL''; a loose and not {\it tight} electron and a {\it tight} muon, denoted ``LT''; and a {\it loose} and not {\it tight} electron and a {\it loose} and not {\it tight} muon, denoted ``LL''. Event categories are denoted ``RR'', ``RF'', ``FR'' and ``FF'', where ``R'' refers to a true prompt electron or muon, and ``F'' refers to a fake or non-prompt electron or muon.

For a given event in a data region, the array $\mathbf{w}$ contains the weights assigned to the event in question and specifies to which category the event
belongs.
This array is made up of four components, denoted  $w_{\mathrm{RR}}$, $w_{\mathrm{FR}}$, $w_{\mathrm{RF}}$ and $w_{\mathrm{FF}}$ and is calculated as:
\begin{equation} 
  \, \left( \begin{array}{c}   w_{\mathrm{RR}} \\ w_{\mathrm{RF}} \\ w_{\mathrm{FR}} \\ w_{\mathrm{FF}} \end{array} \right) 
    = \mathcal{M}^{-1}
     \left( \begin{array}{c} \delta_{\mathrm{TT}} \\ \delta_{\mathrm{TL}} \\ \delta_{\mathrm{LT}} \\ \delta_{\mathrm{LL}} \end{array} \right)~~,
\end{equation}  where \noindent $\delta$ equals unity when the event falls in the given signal or control region, and zero otherwise. 
The matrix $\mathcal{M}$ is written in terms of $r_{e(\mu)}$, the probability for a real {\em loose} electron (muon) to meet the {\em tight} criteria, and $f_{e(\mu)}$, the probability for a fake or non-prompt {\em loose} electron (muon) to meet the {\em tight} criteria, and is calculated as
\begin{equation}
  \mathcal{M} = \left( \begin{array}{cccc}
    r_{e} r_{\mu}            & r_{e} f_{\mu}            & f_{e} r_{\mu}            & f_{e} f_{\mu}        \\
    r_{e} \bar{r}_{\mu}      & r_{e}\bar{f}_{\mu}       & f_{e}\bar{r}_{\mu}       & f_{e}\bar{r}_{\mu}     \\
    \bar{r}_{e} r_{\mu}      & \bar{r}_{e}f_{\mu}       & \bar{f}_{e}r_{\mu}       & \bar{f}_{e} f_{\mu}   \\
    \bar{r}_{e}\bar{r}_{\mu} & \bar{r}_{e}\bar{f}_{\mu} & \bar{f}_{e}\bar{r}_{\mu} & \bar{f}_{e}\bar{f}_{\mu} \\
  \end{array} \right) ~~,
\end{equation}
where \noindent $\bar{x} \equiv 1-x$ for $x=f$ or $r$. Given that the matrix method probabilities, as detailed later, are parameterized as a function of event characteristics such as lepton kinematics and the number of jets, $\mathbf{w}$ is calculated on an event-by-event basis, allowing an improved determination of the background, and therefore the matrix method as described here is a generalization of that presented in Ref.~\cite{Aad:2010ey}.
The estimated background contribution to the signal region due to a given event is given by:
\begin{equation} 
W = r_{e}f_{\mu} \, w_{\mathrm{RF}} + f_{e}r_{\mu} \, w_{\mathrm{FR}} + f_{e}f_{\mu} \, w_{\mathrm{FF}} ~~.
\label{eq:mm}
\end{equation}
The background in a given \metnjets{} bin is given by the sum of $W$ over all events in that bin.
The respective event yields in the opposite-sign and same-sign lepton samples, are shown separately in Table~\ref{t:fakes} 
for the various classes of events used in the matrix method, together with the results, expressed as estimated fake or non-prompt 
background yields in the two samples, integrated over \met{} and \njets{}.

\subsection{Measurement of matrix method probabilities}

The probabilities $r_{\mu}$ for real muons and $r_e$ for real electrons which pass both the loose and tight selection cuts 
are determined with high-purity samples of $Z\to \mu\mu$ and $Z\to ee$ decays, respectively, using a tag and probe method. 

The values of $r_{\mu}$ are measured as a function of muon $\eta$ and jet multiplicity and vary from 0.94 to 0.97. The values of $r_e$ 
are measured as a function of electron $\eta$ and \pt{} for events without jets, and also as a function of the angular distance $\Delta R$
between the electron and nearest jet otherwise.  For events containing two or more jets, $r_e$ is
corrected to better match the expected efficiency in $t\bar{t}$ events.
The correction is calculated from comparisons of $t\bar{t}$ and $Z\to ee$ simulated events.
The complexity of parameterization for the electrons with respect to muons is due to the greater sensitivity of electron identification to 
jet activity. The values of $r_e$ vary from 0.77 to 0.81 from lowest to highest electron \pt{}, from 0.75 to 0.81 from low to high  $|\eta|$, 
and from 0.70 to 0.81 from low to high $\Delta R$ separation between the electron and the nearest jet. Uncertainties on $r_e$ (1\%--2\%) and $r_\mu$ (1\%--4\%) reflect both statistics and variations observed in their determination derived from changes in the modeling of signal and background components in $Z\to ee$ and $Z\to\mu\mu$ 
invariant mass distributions.

The probabilities for jets to be misidentified as muons or for non-prompt muons, $f_\mu$, are measured in a data sample dominated by multijet events selected by requiring low $\met$.
The measurement method employs fits to the transverse impact parameter significance distribution of the candidate muon to disentangle the fake or non-prompt component. 
Over the muon range $|\eta|<2.5$, $f_\mu$ varies from 0.13 to 0.18 and shows less variation with the number of jets, only shifting by about $0.02$ within any particular $\eta$ bin.
An uncertainty on $f_\mu$ is assigned based on the difference with measurements made using an alternative method, in which
specific selection criteria are relied upon to provide a pure sample of muon candidates from fake or non-prompt sources. Measured as a function 
of muon $\eta$ and the \pt{} of the jet with the highest \pt{}, $f_\mu$ 
varies from 0.18 to 0.28. The difference in predicted net background yield from these two $f_\mu$ measurements is taken as the uncertainty on the background estimate, which amounts to about 24\%.

The probabilities for jets to be misidentified as electrons or for non-prompt electrons, $f_e$, are determined in samples dominated by multijet events and parameterized in the same way as $r_e$.
In order to assign a central value and uncertainty for $f_e$, separate criteria are imposed on the multijet events, to enhance the presence of either
fake electrons from jets or electrons from photon conversions in light-flavor quark jets, yielding $f^{\mathrm{jets}}_e \approx 0.15$ and $f^{\mathrm{conv}}_e \approx 0.30$, respectively. 
From data samples enriched in light or heavy quark ($b$ or $c$) jets, it is found that the probability $f_e$ is very similar between the two categories. As the relative composition of fake or non-prompt electrons is not known a priori, a simple average of $f^{\mathrm{jets}}_e$ and  $f^{\mathrm{conv}}_e$ is performed in each \pt{} and $\eta$ bin to give the $f_e$ values.
The uncertainty in each bin is determined as half of the difference between $f^{\mathrm{jets}}_e$ and $f^{\mathrm{conv}}_e$.
In the opposite-sign signal region, the contribution from electrons and muons with mismeasured charge in the inner detector is estimated to be very small and is not accounted for in this analysis. 

\subsection{Validation of background estimate}
The estimate of the background in the signal region was validated using an event sample defined by selection criteria that are the same as those just described, with the exception that a same-sign (SS) $e\mu$ pair is required. Fig.~\ref{SS:njets} shows the jet multiplicity and \met{} distributions in the SS data sample.
This sample is dominated by fake and non-prompt lepton events along with a contribution of prompt leptons from $WZ$ and $ZZ$, and also small contributions from $t\bar{t}W$, $t\bar{t}Z$, and same-sign $W^\pm W^\pm jj$ processes, which are collectively denoted as ``Other prompt bkgd.'' in Fig.~\ref{SS:njets}. 
Opposite-sign events where electron charge is misidentified, predominantly because of bremsstrahlung
in the ID material followed by photon conversion, provide a significant contribution.
This same-sign sample is expected to marginally differ in the exact composition of fake or non-prompt leptons from that of the opposite-sign (OS) sample.
For example, the $W+c$ process preferentially yields a non-prompt lepton with opposite charge to that of the prompt lepton from the $W$ decay. 

A closure test of the matrix method was performed using a collection of simulated samples for processes that could contribute to this background category in the opposite-charge $e\mu$ final state.
This included $W/Z+$jets (including heavy flavor), $W\gamma$+jets, top- or $W$-pair production where at least one of the $W$ bosons decays hadronically, Drell--Yan $\tau$-pair production where one $\tau$ decays hadronically, and $s$- and $t$-channel single top production.
Probabilities were measured using generator-level information in simulated samples of $Z+$jets and multijet production.
The results of calculating the background contribution using the matrix method were compared to those derived from generator-level information and were found to agree within uncertainties.

\begin{figure*}[htbp]
\centering
\begin{tabular}{cc}
\includegraphics[width=0.48 \textwidth]{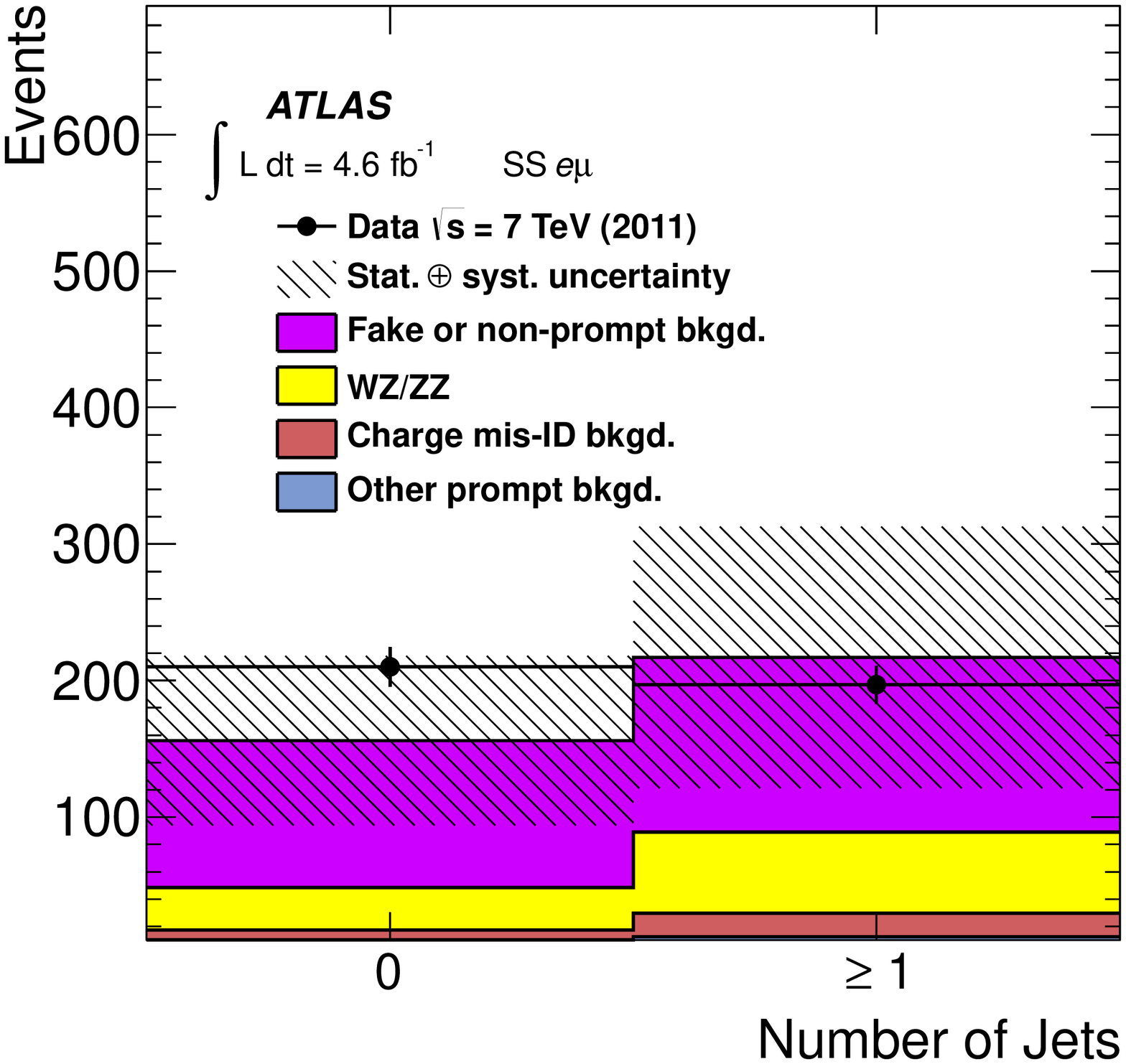} &
\includegraphics[width=0.48 \textwidth]{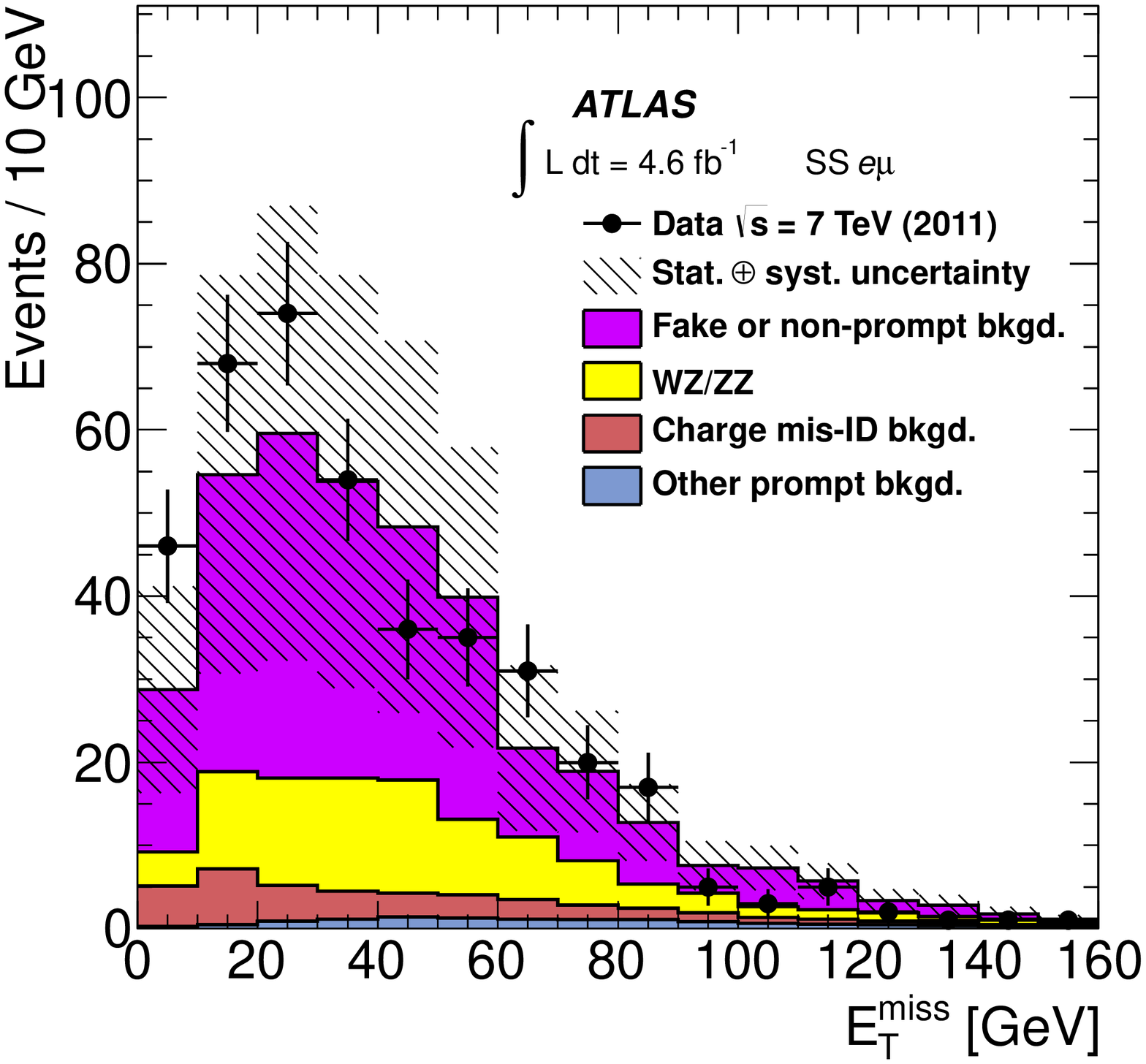} \\
(a) & (b) \\
\end{tabular}
\caption{\label{SS:njets} (a) Jet multiplicity spectrum and (b) missing transverse momentum spectrum for events producing one electron and one muon of same-sign (SS) charge.
The electron and muon candidates and events fulfill the same selection criteria required on the opposite-sign (OS) charge sample.
The hatched regions represent the combination of statistical uncertainty and rate uncertainties on the fake or non-prompt background, as well as uncertainties on the acceptance, efficiency, theoretical cross-sections, and modeling of the processes. The other prompt lepton background category includes contributions from $t\bar{t}W$, $t\bar{t}Z$, and same-sign $W^\pm W^\pm jj$ processes.} 
\end{figure*}

\subsection{Results}
Table~\ref{t:fakes} lists event yields from data in the signal and control regions and the resulting estimation and associated uncertainty of the fake or non-prompt background in both the opposite-sign (OS) and the same-sign (SS) sample.
Signal processes that dominate the OS sample are absent in the SS sample, and the contribution of fake or non-prompt leptons is dominant in the SS event yield as noted previously.
The estimated background in the OS (SS) signal region is $210 \pm 160$ ($240 \pm 120$) events, where the uncertainty is derived from alternative estimates of the background made by varying the electron input probabilities by their associated errors, as well as using muon input probability estimates from the alternative measurement method.
An \njets{} versus \met{} distribution is made for each configuration of matrix method probabilities and later used as input in the likelihood fit in order to assign systematic uncertainties on the signal yields returned in the default fit.

\begin{table}
  \caption{Fake or non-prompt background estimates in the opposite-sign (OS) and same-sign (SS) electron-muon samples.
	Overall data yields are given for the control (LL, LT and TL) and signal (TT) regions, together with the estimates of the backgrounds ($\displaystyle\sum^{\mathrm{Bins}}\sum^{\mathrm{Events}}W$).
    The events from the control regions are used to produce the background estimate after applying the appropriate weights from Eqn.~\ref{eq:mm}.
    Backgrounds are shown with their statistical, electron-related, and muon-related systematic uncertainties.}
    \label{t:fakes}
    \begin{ruledtabular}
  \begin{tabular}{lcc}
    Region & \multicolumn{2}{c}{Event yields}  \\  
           & OS & SS  \\  \hline
    LL     &  $3560$ & $1623$ \\   
    LT     &  $4744$ &  $896$ \\   
    TL     &  $1137$ &  $499$ \\ \hline  
    TT     & $12224$ &  $407$ \\ \hline
    &  \multicolumn{2}{c}{Estimated fake or non-prompt background} \\
    $\displaystyle\sum^{\mathrm{Bins}}\sum^{\mathrm{Events}}W$ & $210 \pm 20 \pm 150 \pm 50$ & $240 \pm 10 \pm 120 \pm 10$ \\
   \end{tabular}
\end{ruledtabular}
\end{table}

\section{Fit method and uncertainties}
\label{Sec:FitMeth}
Templates in the \metnjets{} parameter space are produced for signal processes (\ttbar{}, \WW{}, \Ztau{}) and backgrounds (\Wt{}, \WZ{}/\ZZ{}, fake and non-prompt) by applying the object and event selection described above.
These templates are employed in a fit to data.
The parameter space is divided into two bins of jet multiplicity, $\njets{}= 0$ and $\njets{}\geq 1$, counting reconstructed jets with $\pt \geq 30~\GeV$.
The \met{} distribution is divided into twenty bins from $0 <\met{}< 200\GeV$ in increments of $10\GeV$, with the bins bordering $200\GeV$ also containing the overflow of events with $\met{} \geq 200\GeV$.
Studies using simulated samples found the choices of two jet multiplicity bins and of a jet threshold $\pt{} \geq 30\GeV$ to be optimal in terms of minimizing statistical and systematic uncertainties, such as those arising from jet energy scale effects and \ttbar{} modeling.

Normalized templates for signal and background components are used to construct a binned likelihood function that is maximized in the fit to data. The normalization parameters of the \ttbar{}, \WW{} and \Ztau{} templates are treated as free parameters in the fit, whereas the normalization parameters of the \Wt{} and \WZ{}/\ZZ{} templates are constrained to their expected values. The template for background involving at least one fake or non-prompt lepton candidate is constrained to the estimate derived from data as described previously in Sec.~\ref{Sec:DDB}. The templates for \ttbar{} and \WW{} include electrons and muons from tau-lepton decays. 

The fiducial region in this analysis is defined by particle level quantities chosen to be similar to the selection criteria used in the fully reconstructed sample. Electrons must have transverse energy $E_{\mathrm{T}} > 25$~\GeV{} and pseudorapidity $|\eta| < 2.47$, excluding the transition region $1.37 < |\eta| < 1.52$.
Muons are required to have transverse momentum $\pt > 20$~\GeV{} and pseudorapidity $|\eta| < 2.5$.
All selected electron or muon particles must originate from a $W$ boson decay from the hard scattering process, or from tau-lepton decays that themselves are 
from a $W$ boson or $Z$ boson decay. A further correction applied to leptons, to include the momenta contribution of photons from narrow-angle QED FSR, is the addition of the
momenta of all photons within a cone of $\Delta R = 0.1$ around the lepton to its momentum. 

Fitted event yields are used to extract fiducial and full cross-sections for the signal processes. The former is desirable because it is 
a quantity that is closer to what is measured by the detector and does not suffer from theoretical extrapolation errors. The
two cross-sections are calculated as:
\begin{equation}
\label{eq:fid_xsection}
\sigma^{\mathrm{fid}}_X =   \frac{\Nfitfid}{\C  \cdot \lum}, 
\end{equation}
\begin{equation}
\label{eq:tot_xsection}
\sigma^{\mathrm{tot}}_X =   \frac{\Nfittot}{\A \cdot \C  \cdot  B(X \ra \emu+Y) \cdot \lum}
\end{equation} 
respectively, where \lum{} corresponds to the integrated luminosity of the data sample; $\A$ is the kinematic and geometric acceptance of the fiducial region as a fraction of the complete phase space; $\C$ is the ratio of the number of events fulfilling the offline selection criteria to the number of events produced in the fiducial region estimated from simulation; \Nfittot{} (\Nfitfid{}) is the number of events attributed to the specified process by the fit using systematic uncertainties that affect $\A \cdot \C$ ($\C$ only); and $B(X \ra \emu+Y)$ is the branching fraction to inclusive \emu{} final states for the decay channel under consideration taking into account the branching fractions of tau-lepton decays to electrons and muons.  

Systematic uncertainties are estimated by examining their effects on the nominal templates.
These effects are broadly broken up into two categories, those affecting normalization and those affecting the shape of predicted templates, which are calculated using Monte Carlo pseudo-experiments.
Each source of uncertainty considered may affect both template normalization and shape, with the exception of integrated luminosity and LHC beam energy uncertainties, which affect only template normalization.
Uncertainties associated with the fake or non-prompt background and parton distribution function modeling are handled differently as special cases, described detail below.
The dominant sources of systematic uncertainties are listed in Table~\ref{t:sys_error} for the signal processes.
For background templates, most of the uncertainties listed in Table~\ref{t:sys_error} are applied with the exception of Monte Carlo model uncertainties, LHC beam energy, and PDF uncertainties.

\subsection{Template normalization uncertainties}
Systematic uncertainties affecting the acceptance, efficiency and background cross-sections are incorporated as Gaussian constrained parameters in
the likelihood function. The Gaussian probability distributions for each systematic
uncertainty parameter multiply the likelihood thus profiling the uncertainty. These terms penalize the likelihood if the parameters move
away from their nominal values during the minimization procedure.

\subsection{Template shape uncertainties}
Monte Carlo ``pseudo-experiments'' are performed to estimate uncertainties on event yields due to systematic uncertainties affecting template shapes.
For a given source of systematic uncertainty, $S$, sets of modified \metnjets{} signal and background templates are produced in which $S$ is varied up and down by its expected uncertainty, while the template normalization remains fixed to its assumed standard model expectation.
Pseudo-experiments are performed by fitting these modified templates to ``pseudo-data'' randomly drawn according to the nominal (\ie{}, no systematic effects applied) templates.

Pseudo-data are constructed for each pseudo-experiment using the expected 
number of events, $\bar{N}_X$, and \metnjets{} shape for each process $X$. 
For each pseudo-experiment the following procedure is carried out.
The expected number of events for process $X$ is sampled from a Gaussian distribution 
of mean $\bar{N}_X$ and width determined by the uncertainty on $\bar{N}_X$.
This number is then Poisson fluctuated to determine the number of events, $N_X$, 
for process $X$. 
The shape of process $X$ in the \metnjets{} parameter space is then 
used to define a probability distribution function from which to sample the $N_X$
events contributing to the pseudo-data for the pseudo-experiment.
This is repeated for all processes to construct the pseudo-data in the 
 \metnjets{} parameter space as the input to the pseudo-experiment.
The pseudo-experiment is then performed by fitting the pseudo-data to the
modified templates and extracting the number of events for each signal process, 
$N_{\mathrm{sig}}$. This procedure is repeated one thousand times to obtain a well-defined
distribution of $N_{\mathrm{sig}}$ values.

The difference, $\Delta N_{\mathrm{sig}}$, between the mean value of this distribution and $\bar{N}_X$ is taken as the error due to template shape effects.
To obtain the final template shape uncertainty, each positive $\Delta N_{\mathrm{sig}} / N_{\mathrm{sig}}$ value is added in quadrature to obtain the total positive error, and each negative value is added likewise to obtain the total negative error.

\subsection{Fake or non-prompt background uncertainties}
To evaluate the uncertainty on the fake or non-prompt background contribution, the matrix method input probabilities are varied; the background templates are then re-derived and the measurement is repeated.
The observed maximum deviation of the signal parameters measured from templates where electron probabilities are varied is assigned as an uncertainty.
Similarly the deviation observed when using the alternative set of muon probabilities is assigned as an uncertainty.
The net uncertainty is calculated as a quadratic sum of both uncertainties. 

\subsection{PDF uncertainties}
The uncertainties associated with the choice of parton distribution functions are evaluated using a number of different PDF sets.
The envelope of uncertainty bands from the CT10~\cite{CT10PDF}, MSTW2008~\cite{MSTW2008} and NNPDF~2.3~\cite{Ball:2012cx} sets is determined using the procedure prescribed for LHC studies~\cite{PDF4LHC}.
There are two PDF-related uncertainties defined, which are the \emph{Intra-PDF uncertainty} and the \emph{Inter-PDF uncertainty}.
The former is the uncertainty within a given PDF set originating from uncertainties on various inputs to the PDF calculation or other uncertainties assigned by the particular PDF set authors.
The latter is the variation observed when comparing one PDF to another.
The comparison is made using the central value of each PDF set and measuring the variation of the observable.
The full PDF set uncertainty combines the inter- and intra-PDF uncertainties by taking the envelope of the minimum and maximum of these values. Uncertainties associated with the parton distribution functions are not profiled in the fit.
Shape uncertainties are measured by fitting the varied templates to data while variations between calculated \A{} and \C{} values are used to assign acceptance uncertainties. Fitting the templates with different PDF sets to data results in yield uncertainties, the envelope of which is taken as the PDF shape uncertainty.
The PDF set uncertainties, shown in Table~\ref{t:sys_error}, are computed in this way to avoid the complexity that would otherwise be introduced into the fit
if they were to be profiled.

\subsection{LHC luminosity and beam energy}
The uncertainty in the measured integrated luminosity is 1.8\%, which affects both the fitted yields and the calculated cross-sections for signal and background templates, while the uncertainty associated with the center-of-mass collision energy, $\sqrt{s}$, affects the production cross-sections.
The beam energy can be calibrated using the revolution frequency (RF) difference between protons and lead ions.
The RF is different for lead ions and protons due to their different ratio of charge to rest mass, and depends on the LHC dipole field setting.
The calibration can be performed because the proton beam momentum is proportional to the square root of the proton's RF divided by the frequency difference~\cite{Wenninger:2013jw}. 
The nominal beam energy at $\sqrt{s}= 8\,$ \TeV{} was calibrated to be $3988 \pm 5 \pm 26$ GeV during $p+$Pb runs in early 2013~\cite{Wenninger:2013jw}
and corresponds to a relative uncertainty of $0.66\%$, which is assumed to be the same for $\sqrt{s}=7$ \TeV{}. 
Both of these sources of uncertainty affect template normalization but have no effect on template shape, unlike other uncertainties which affect both normalization and shape.

\subsection{Summary of systematic uncertainties}
{\squeezetable
\begin{table*}
  \caption{Summary of dominant systematic uncertainties.
    Uncertainties expressed as a percentage are shown for each signal process, broken down into normalization effects on \C{} (
    the factor relating the measured events to the fiducial phase-space) and $\A \cdot \C$ (the factor relating the measured events 
    to the full phase-space), and template shape effects.
    The normalization uncertainties on $\A \cdot \C$ and $\mathcal{C}$ are symmetrized.
    The reconstruction uncertainties are applied to $\mathcal{C}$ and affect both the fiducial and full cross-section measurements.
    The theoretical uncertainties due to template shape are applied to both the fiducial and full cross-section measurements as well.
    Uncertainties on the fake and non-prompt background, luminosity, and LHC beam energy, which are not divided into normalization and shape components, are listed together.
    \label{t:sys_error}}
    \begin{ruledtabular}
\renewcommand{\arraystretch}{1.2}
  \begin{tabular}{lccccccccc}
    &  \multicolumn{9}{c}{Systematic Uncertainties (\%)}  \\          
     \multirow{2}{*}{\backslashbox{Source}{Process}} &  \multicolumn{3}{c}{$t\bar{t}$} &  \multicolumn{3}{c}{$WW$} & \multicolumn{3}{c}{$\Ztau$}\\           
       &  $\C$ & $\A \cdot \C$  &  Shape & $\C$ & $\A \cdot \C$   &  Shape & $\C$ & $\A \cdot \C$   &  Shape  \\ \hline 
    ISR/FSR+Scale    
    & $\pm 1.1$ & $\pm 0.4$ & $+1.0(-1.5)$ 
    & $\pm 1.0$ & $\pm 0.8$ & $+4.7(-3.5)$ 
    & $\pm 1.1$ & $\pm 0.4$ & $+0.7(-1.0)$ \\ 
    Generator        
    & $\pm 0.7$ & $\pm 0.8$ & $+0.2(-0.0)$ 
    & $\pm 0.6$ & $\pm 0.5$ & $+4.5(-0.4)$ 
    &                         & &$+0.0(-0.7)$ \\	    
    PS Modeling     
    & $\pm 0.9$ & $\pm 0.6$ & $+0.0(-0.1)$ 
    & $\pm 0.5$ & $\pm 1.0$ & $+3.5(-0.0)$ 
    &   & & $+0.0(-0.6)$ \\ 
    $\Ztau$ PS Modeling  
    &  & & $+0.0(-0.5)$
    &  & & $+0.0(-0.6)$ 
    & $\pm 1.8$ & $\pm 3.3$ &  $+0.5(-0.0)$ \\ 
    PDF                          
    & $\pm 0.6$ & $\pm 1.7$ & $\pm 0.5$ 
    & $\pm 0.1$ & $\pm 0.7$ & $\pm 1.6$ 
    & $\pm 0.2$ & $\pm 1.3$ &  $\pm 0.8$ \\ \hline 
    $e$ reco., ID, isolation    & $\pm 3.2$ & &$+0.0(-0.1)$ & $\pm 3.2$ & &$+0.3(-0.3)$ & $\pm 3.3$ &  &$+0.0(-0.8)$ \\       
    $\mu$ reconstruction        & $\pm 0.8$ & &$+0.0(-0.0)$ & $\pm 0.8$ & &$+0.0(-0.0)$ & $\pm 0.8$ &  &$+0.0(-0.0)$ \\ 	    	    
    \met{}--cellout & $\pm 0.0$ & &$+0.4(-0.2)$ & $\pm 0.0$ & &$+8.1(-9.9)$ & $\pm 0.0$ &  &$+2.3(-0.2)$ \\        
    \met{} pile-up  & $\pm 0.0$ & &$+0.1(-0.1)$ & $\pm 0.0$ & &$+3.7(-4.5)$ & $\pm 0.0$ &  &$+1.0(-1.7)$ \\   	            
    Jet energy scale       & $\pm 0.8$ & & $+1.4(-1.4)$  & $\pm 0.6$ & & $+0.5(-4.8)$      & $\pm 0.5$ &  & $+1.4(-3.1)$               \\
    Jet energy resolution       & $\pm 0.2$ & &$+0.3(-0.0)$ & $\pm 0.2$ & &$+0.0(-2.6)$ & $\pm 0.2$ &  &$+0.0(-0.1)$ \\ 	    
    Jet vertex fraction         & $\pm 0.8$ & &$+0.1(-0.0)$ & $\pm 0.3$ & &$+0.0(-1.7)$ & $\pm 0.2$ &  &$+0.0(-0.3)$ \\   \hline 		    
                                &  \multicolumn{3}{c}{$t\bar{t}$} &  \multicolumn{3}{c}{$WW$} & \multicolumn{3}{c}{$\Ztau$}    \\  \hline          
    Fake or non-prompt background        & \multicolumn{3}{c}{$\pm 0.8$} & \multicolumn{3}{c}{$\pm 5.6$} & \multicolumn{3}{c}{$\pm 0.7$} \\ 
    Luminosity                           & \multicolumn{3}{c}{$\pm 1.8$} & \multicolumn{3}{c}{$\pm 1.8$} & \multicolumn{3}{c}{$\pm 1.8$} \\ 
    LHC beam energy                      & \multicolumn{3}{c}{$\pm 1.8$} & \multicolumn{3}{c}{$\pm 1.0$} & \multicolumn{3}{c}{$\pm 0.8$} \\ 
  \end{tabular}
\end{ruledtabular}
\end{table*}
}

Table~\ref{t:sys_error} lists the sources and effects of the most significant systematic variations on the acceptance correction factors and on the event yields derived from the fit. The first group of entries in the table are the theoretical uncertainties. To determine the uncertainty due to the choice 
in the modeling of a particular aspect of the event, comparisons are made between Monte Carlo samples featuring alternative choices to the default ones. The
uncertainty on the modeling of additional QCD radiation on \ttbar{} and $WW$ is evaluated by comparing \mcnlo{} to \alpgen{} where the default scales are varied simultaneously by factors
 of 2 and 0.5.
 The uncertainty due to the choice of Monte Carlo generator is determined for \ttbar{} and $WW$ 
 by comparing the default generators to \powheg{} while the uncertainty due to the modeling of the parton shower and fragmentation
 is evaluated by interfacing the default generators to \pythia{}.  In the case of \Ztau{}, the theoretical uncertainties
are calculated by comparing \sherpa{} to the appropriate \alpgen{} sample interfaced to \herwig{}. The evaluation of the uncertainty
due to the choice of PDF has been described Sec.~\ref{Sec:FitMeth}~D.

The second group of entries in Table~\ref{t:sys_error} correspond to the experimental uncertainties. 
The uncertainties associated with Monte Carlo modeling of the lepton trigger, reconstruction and identification efficiencies are evaluated by studying
${Z \to ee/Z \to \mu\mu}$ and ${W \to e\nu/W \to \mu\nu}$ events selected from data as well as ${Z \to ee/Z \to \mu\mu}$, ${W \to e\nu/W \to \mu\nu}$, 
and \ttbar{} events from simulation~\cite{Aad:2014fxa}. The dominant experimental uncertainties on template normalization stem from electron reconstruction, 
identification, and isolation. These uncertainties are large due to the difference in efficiency of the isolation cut between the $Z+$jets region 
where the efficiency is measured and the rest of the signal region.

The main contributors to the uncertainty on \met{}
originate from calorimeter cells not associated with
any physics object (\met{}--cellout term) and the pile-up correction factors. In fact
the former is responsible for the single largest contribution and
results, in the $WW$ measurement, in shape uncertainties in excess of $10\%$ which is a dominant source of uncertainty on the full and 
fiducial cross-section values. 

The uncertainty on the jet energy scale also leads to relatively large template shape uncertainties for all signal processes.
In the central region of the detector ($|\eta|<1.7$) the jet energy scale uncertainty varies from 2.5\% to 8\% as a function of jet \pt{} and $\eta$~\cite{ATLAS-CONF-2011-032}, as estimated from {\em in situ} measurements of the detector response.
This uncertainty estimate includes uncertainties from jet energy scale calibration, calorimeter response, detector simulation, and 
the modeling of the fragmentation and UE, as well as other choices in the Monte Carlo event generation.
Intercalibration of forward region detector response from the central regions of the detector also contributes to the total uncertainty on jet energy scale.
Additional uncertainties due to pile-up and close-by jet effects are also included.
The uncertainty introduces distortions in the template shapes including effects propagated to the calculation of \met{}.
To obtain an estimate of this source of uncertainty, the jet energy scale is broken into sixteen independent components.
Each component is individually shifted up and down within its uncertainties for a total of 32 variations in the evaluation of shape uncertainties, the results of which are combined and shown as a single entry in Table~\ref{t:sys_error}.

The jet energy resolution has been found to be well modeled by simulation. It is measured
from calorimeter observables by exploiting the transverse momentum balance in events containing jets with large \pt{}. Two independent
in situ methods sensitive to different sources of systematic uncertainties are used to measure the resolution
which the Monte Carlo simulation describes within 10\% for jets whose \pt{} ranges from $30- 500$ GeV~\cite{Aad:2012ag}. 
The uncertainty due to the JVF is determined from studies of $Z \to ee/\mu\mu+$jets events. 

The last group of entries on Table~\ref{t:sys_error} includes uncertainties on fake or non-prompt backgrounds, the measurement of integrated luminosity, and the determination of the LHC beam energy.
The uncertainty due to modeling of the fake or non-prompt background, whose evaluation is described
in Sec.~\ref{Sec:FitMeth}~C, has the greatest effect on the $WW$ measurement. The uncertainty in the integrated luminosity 
is dominated by the accuracy of the beam separation-scans and the resulting uncertainty of 1.8\% is assigned to each signal process. 
The uncertainty of 0.66\% on the beam energy is found to vary
the prediction for \ttbar{} production, calculated at NNLO plus next-to-next-to-leading logarithm by \toppp{}~\cite{Top++2011}, by $1.8\%$. Similarly,
for $WW$ and \Ztau{}, an equivalent study was performed with predictions at NLO from \mcfm{} v6.6~\cite{MCFM}, resulting in variations of $1.0\%$ and $0.8\%$ respectively. These variations are assigned as uncertainties to the measured cross-sections as shown in the last item of Table~\ref{t:sys_error}.

Overall since the $WW$ and \Ztau{} signals overlap in the 0-jet bins, most of the significant shape uncertainties involve the wrong assignment of events to one of these two samples.
Very few effects can move a $WW$ or \Ztau{} event into the $\ge 1$ jet bin, so generally small shape uncertainties on \ttbar{} are observed, where interference from the other processes is minimal.
This event assignment uncertainty affects $WW$ approximately three times more than \Ztau{} due to the larger yield of \Ztau{} events.

The main contributions to the uncertainty on $\A \cdot \C$, as shown in Table~\ref{t:sys_error}, are
the PDF for \ttbar{} and the PS modeling for $WW$ and \Ztau{}.
The theoretical uncertainties on the correction factors \C{} are small.
No individual source of theoretical uncertainty on \C{} exceeds the uncertainty due to experimental effects (dominated by 
those associated with electron scale factors and luminosity).
One effect observed from this table is that there is apparent anti-correlation between uncertainties on \A{} and \C{}, leading to an uncertainty on their product that is smaller than that on the multiplicands, e.g. the ISR/FSR+scale uncertainty.
Uncertainties on branching ratios~\cite{Beringer:1900zz} used in the cross-section calculations
are negligible relative to experimental uncertainties and not included in Table~\ref{t:sys_error}.  

Within the fiducial region, uncertainties on \C{} come mainly from experimental sources and template shape uncertainties.
The dominant source varies between signals; template shape uncertainties are dominant in the $WW$ measurement, where the likelihood fit is sensitive to variation in the scale of \met{}--cellout terms.
The uncertainty on the fiducial \ttbar{} cross-section is dominated by the electron reconstruction, identification and isolation.
In the \Ztau{} channel, leading uncertainties derive from PS modeling and the jet energy scale measurement. 
\section{Results}
\label{Sec:Xsec}

\subsection{Event yields}
Comparisons between data and Monte Carlo predictions together with event yields before the application of the fitting procedure are displayed in Fig.~\ref{fig:dist_os_prefit} and Table~\ref{t:yield_os}.
The Monte Carlo predictions are normalized to the values given in Sec.~\ref{Sec:MC}.
These comparisons are shown in the signal region and sub-divisions thereof based on jet multiplicity calculated for jets above the 30~\GeV{} \pt{} threshold and on events with reconstructed \met{} below and above 30~\GeV.
The events shown here satisfy the OS and tight identification criteria specified in Sec.~\ref{Sec:ObjEvtSelec}.
The inclusive yields represent the sum of the binned yields in the \metnjets{} parameter space, which provide the templates used in the fit to the data.
The data yield is observed to be in good overall agreement with the prediction.

The same comparisons are shown after the fitting procedure in Fig.~\ref{fig:dist_os_postfit} and Table~\ref{t:yield_os_fitted} for the signal region and for sub-divisions thereof, based on the classification defined above.
In Fig.~\ref{fig:dist_os_postfit} the error bands are smaller in general than in Fig.~\ref{fig:dist_os_prefit} since they do not include the uncertainties on the theoretical cross-sections for the three signal processes that are included in the pre-fit results.
As expected, yields for the signal processes given by the fit rise with respect to the pre-fit normalization to better fit the observed yield in data.
Furthermore, good agreement is observed within each of the categories shown in Table~\ref{t:yield_os_fitted}, indicating that the background estimation and signal template shapes provide a good description of the data.

\begin{figure*}[p]
  \begin{center}
    \begin{tabular}{cc}
      \includegraphics[width=0.48\textwidth]{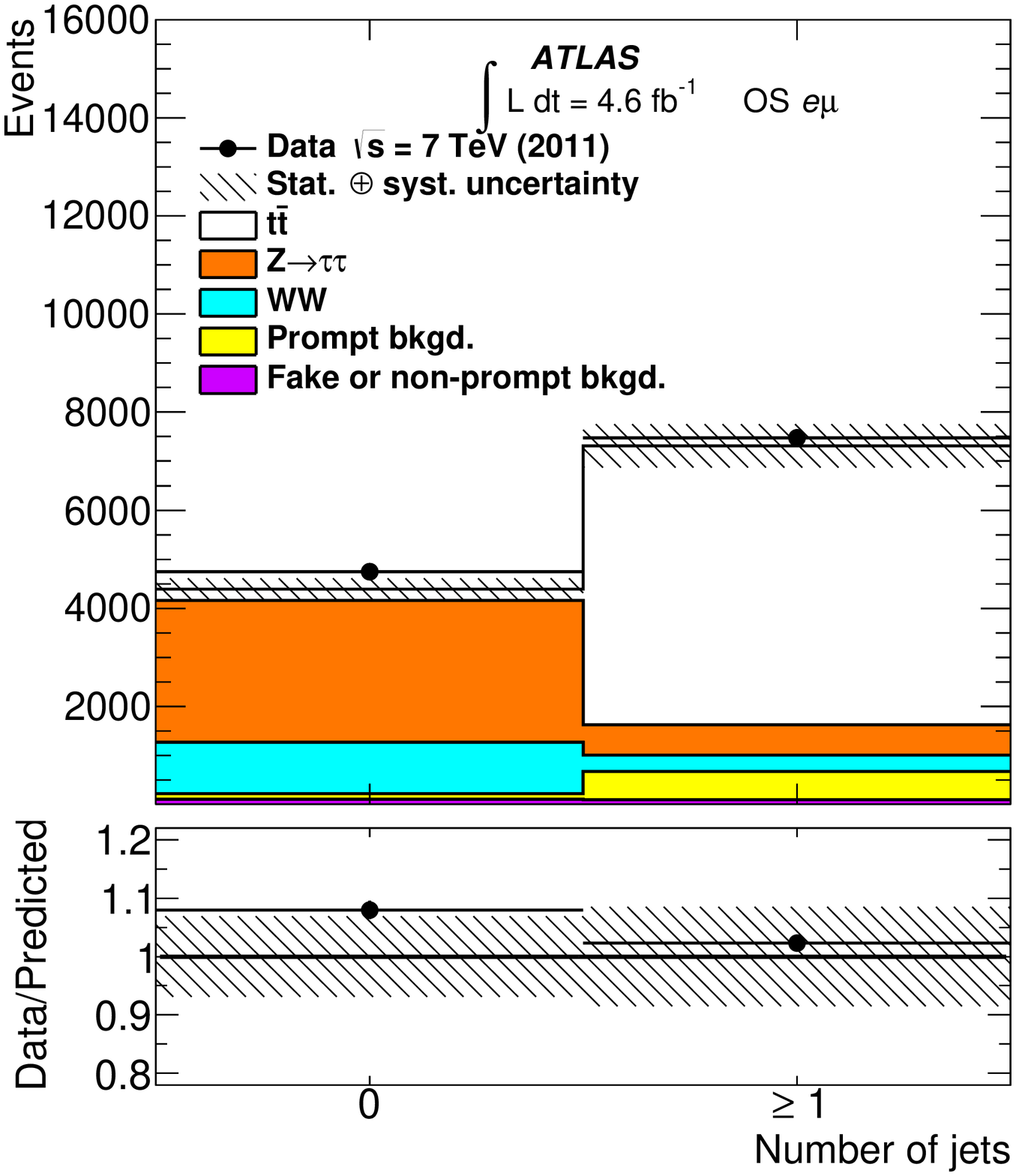} &
      \includegraphics[width=0.48\textwidth]{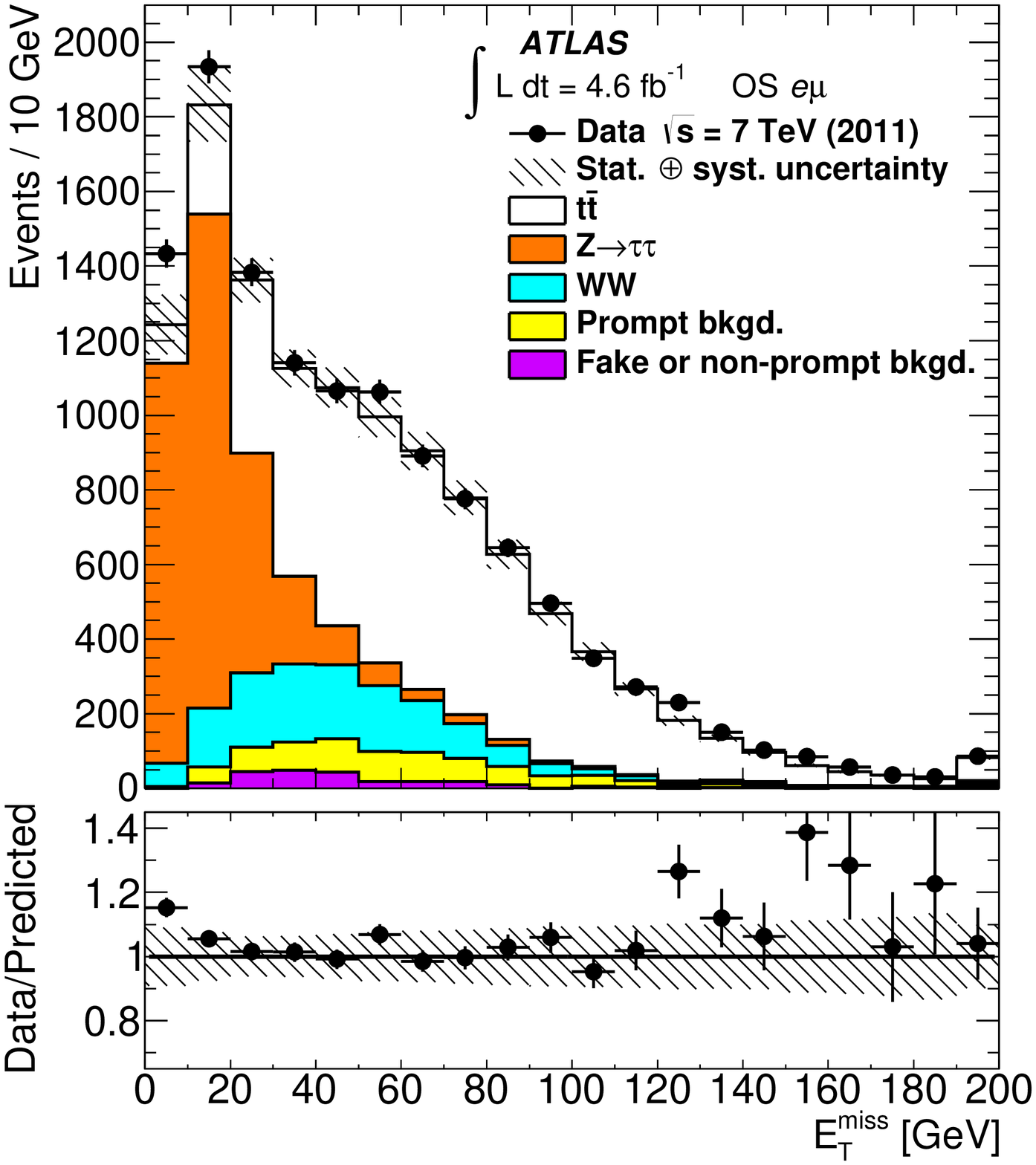}  \\
      (a) & (b) \\
      \includegraphics[width=0.48 \textwidth]{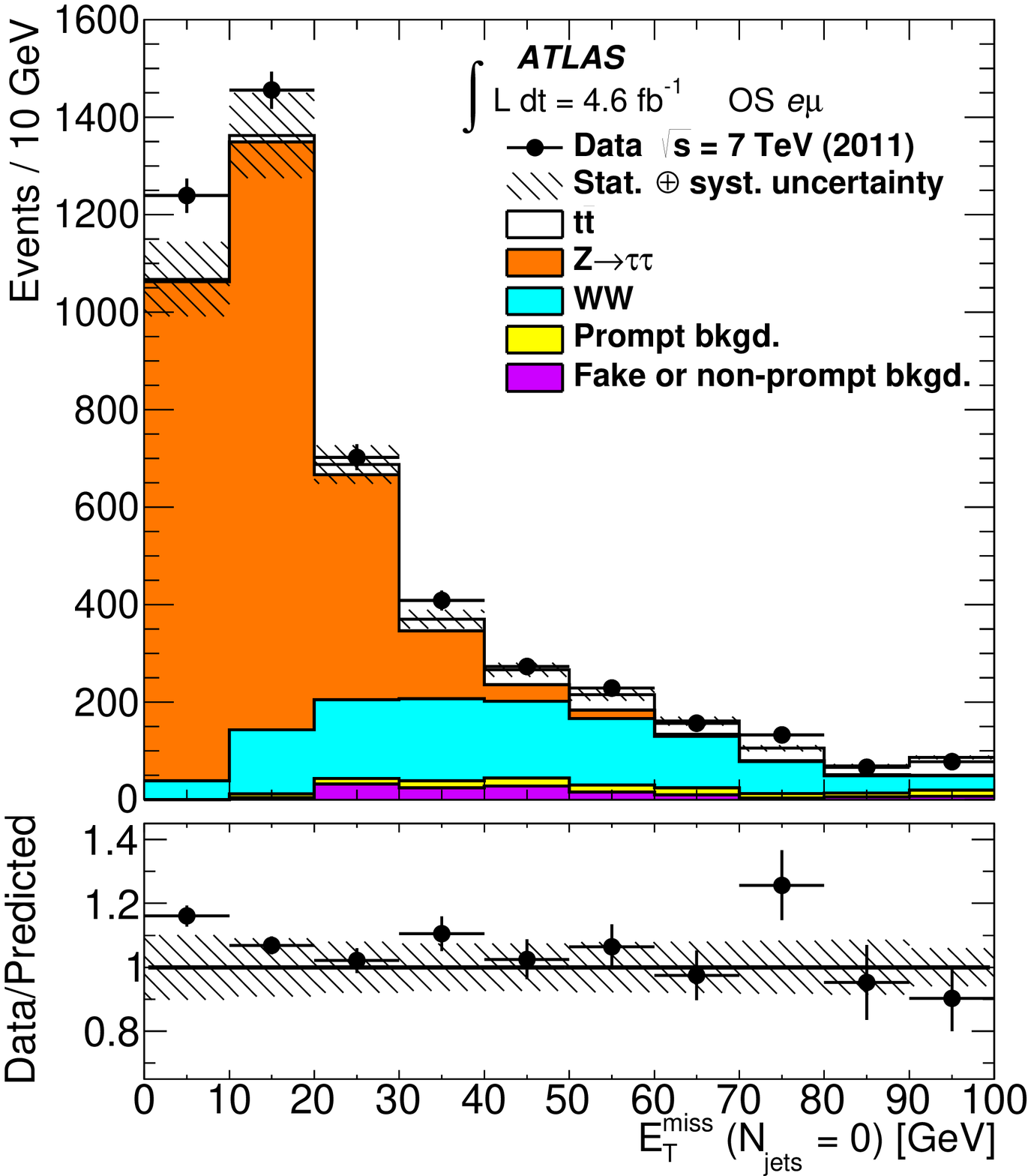} &
      \includegraphics[width=0.48 \textwidth]{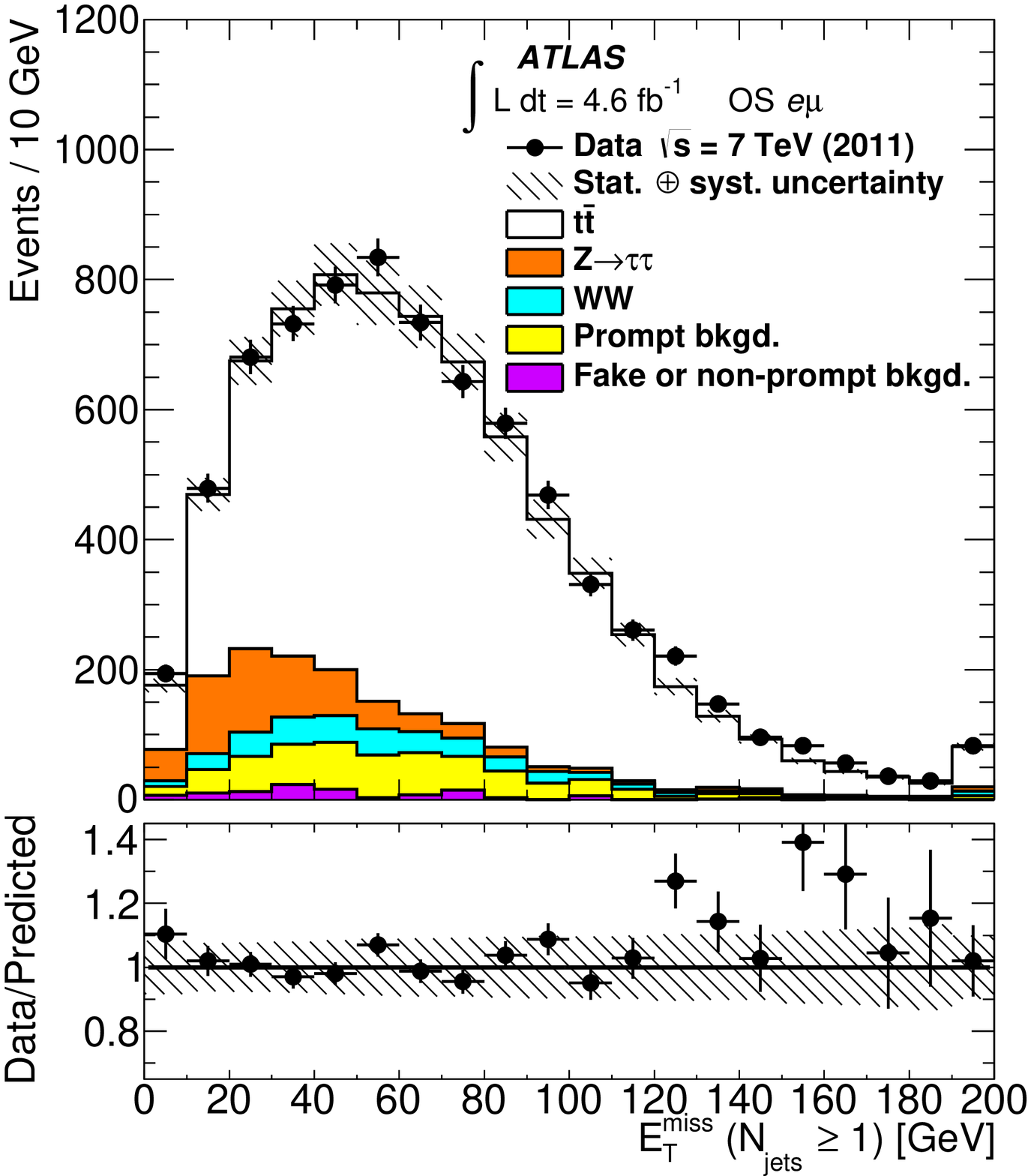} \\
      (c) & (d) \\
    \end{tabular}
  \end{center}
  \caption{Comparison between data and Monte Carlo samples (including the data-driven fake or non-prompt background) normalized to their theoretical cross-sections
  for an integrated luminosity of \lumi{} for events producing one electron and one muon of OS charge: (a) \njets{}, with bins corresponding to 0-jets and $\geq 1$-jet; (b) missing transverse momentum spectrum, \met{}; 
    (c) \met{} for $N_{\rm jets} = 0$  and (d) \met{} for $N_{\rm jets} \geq 1$.
    The electron and muon satisfy the signal region selection criteria presented in Sec.~\ref{Sec:ObjEvtSelec}.
    The hatched regions represent the combination of statistical and systematic uncertainties as described in Table~\ref{t:sys_error} (except for shape uncertainties) together with the full theoretical cross-section uncertainties for the \ttbar{}, \WW{}, and \Ztau{} signal processes.
    The last bins in (b), (c) and (d) contain overflow events. \label{fig:dist_os_prefit}}
\end{figure*}

\begin{figure*}[p]
  \begin{center}
    \begin{tabular}{cc}
      \includegraphics[width=0.48\textwidth]{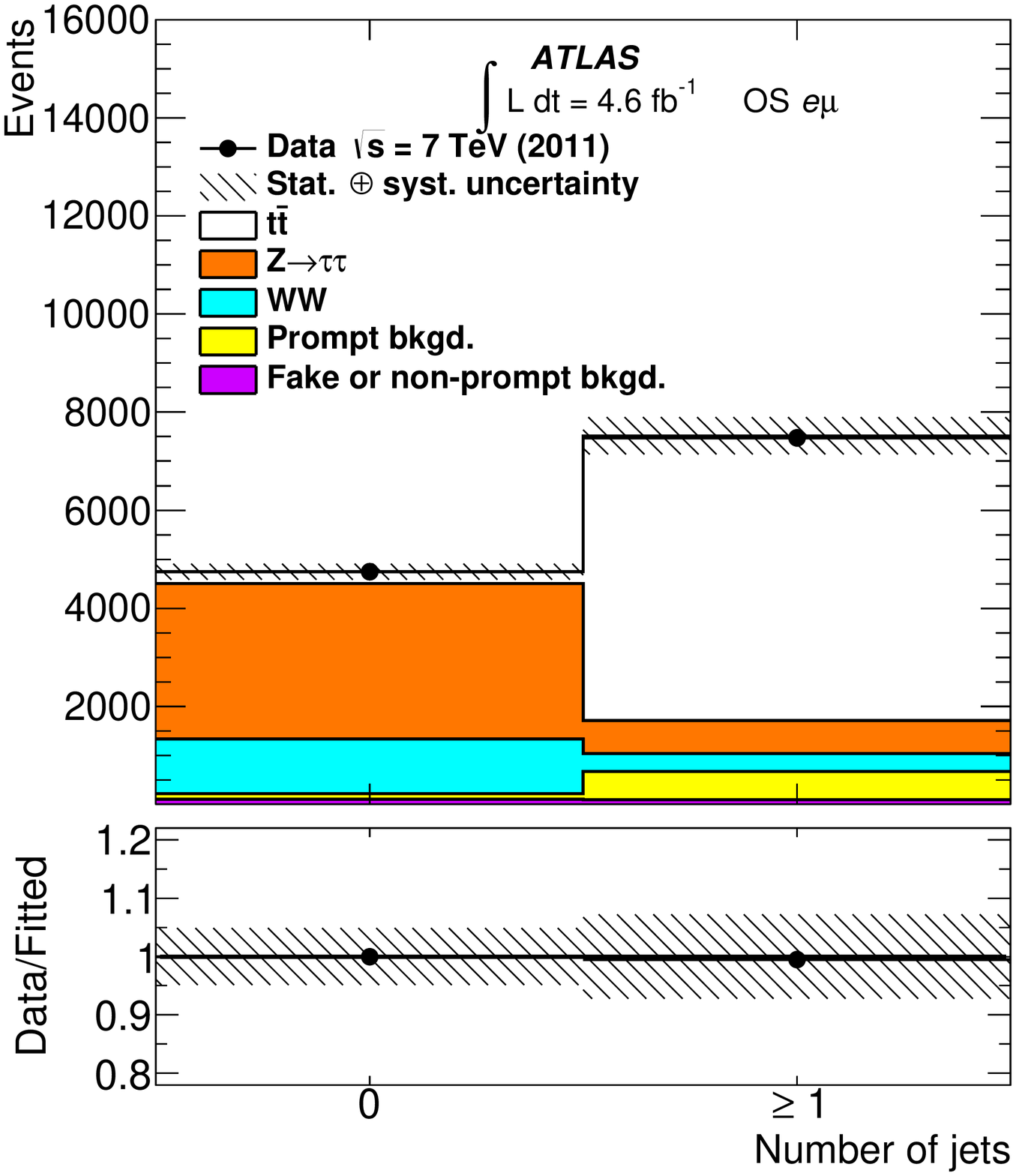} &
      \includegraphics[width=0.48\textwidth]{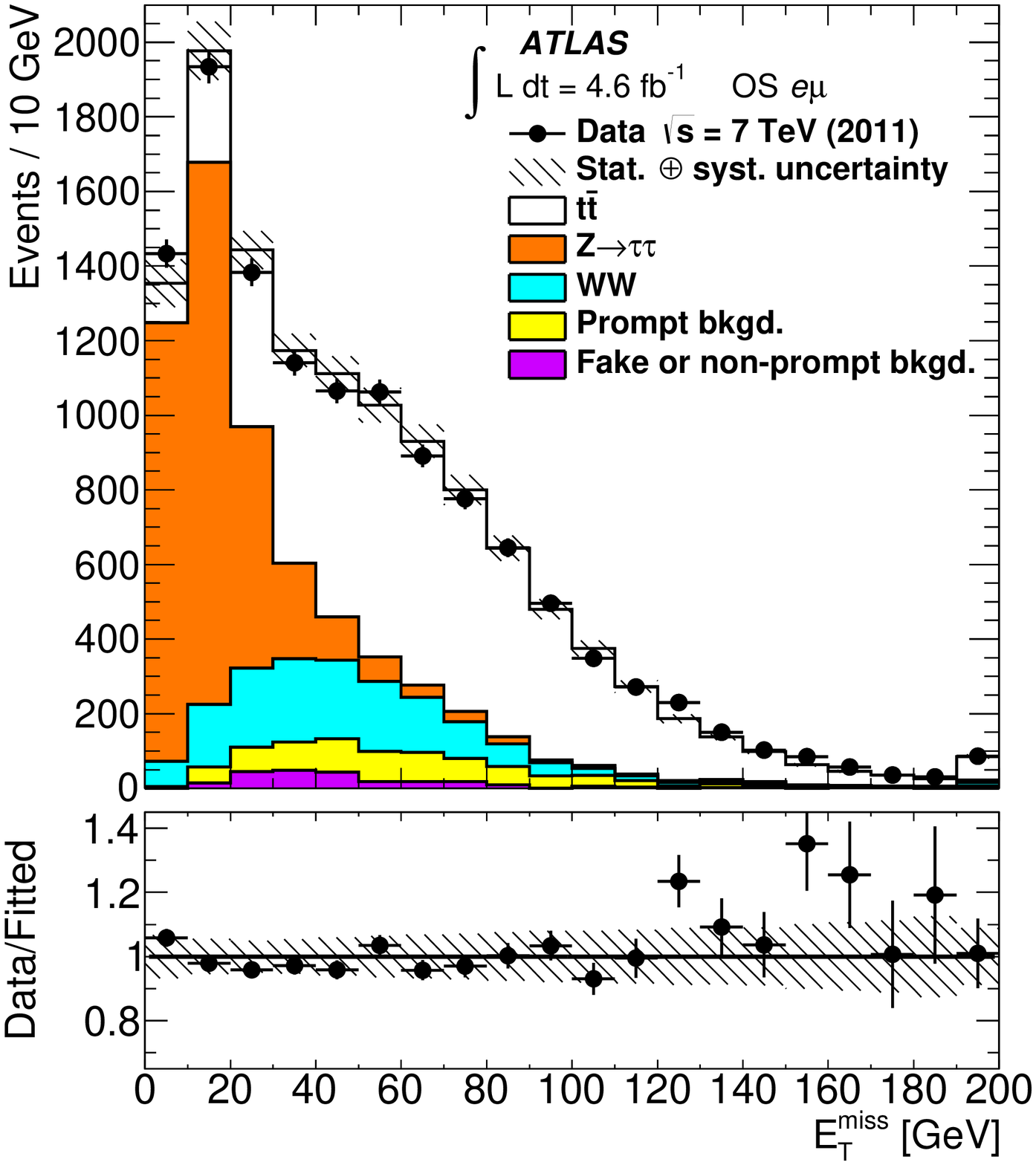} \\
      (a) & (b) \\
      \includegraphics[width=0.48 \textwidth]{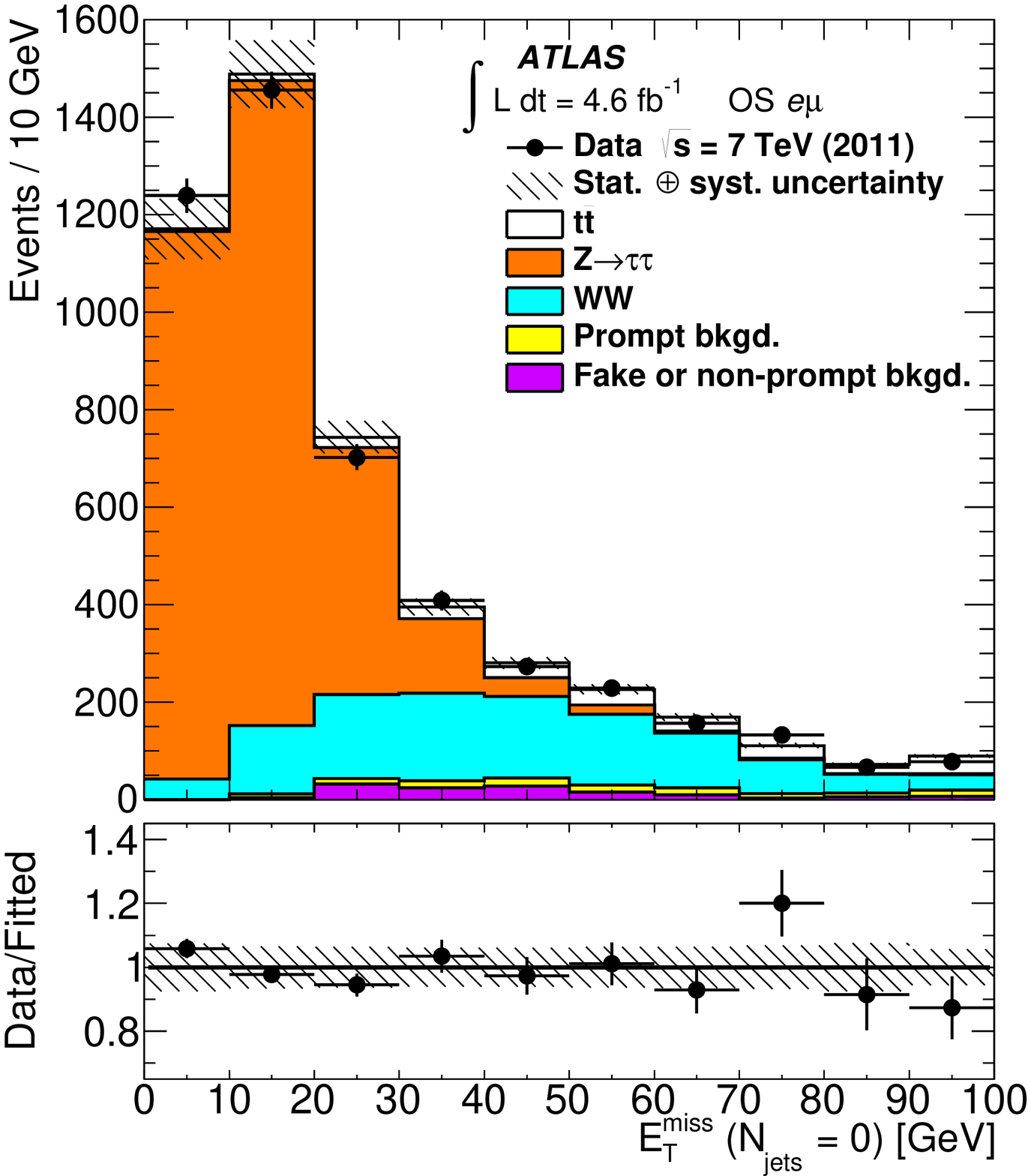} &
      \includegraphics[width=0.48 \textwidth]{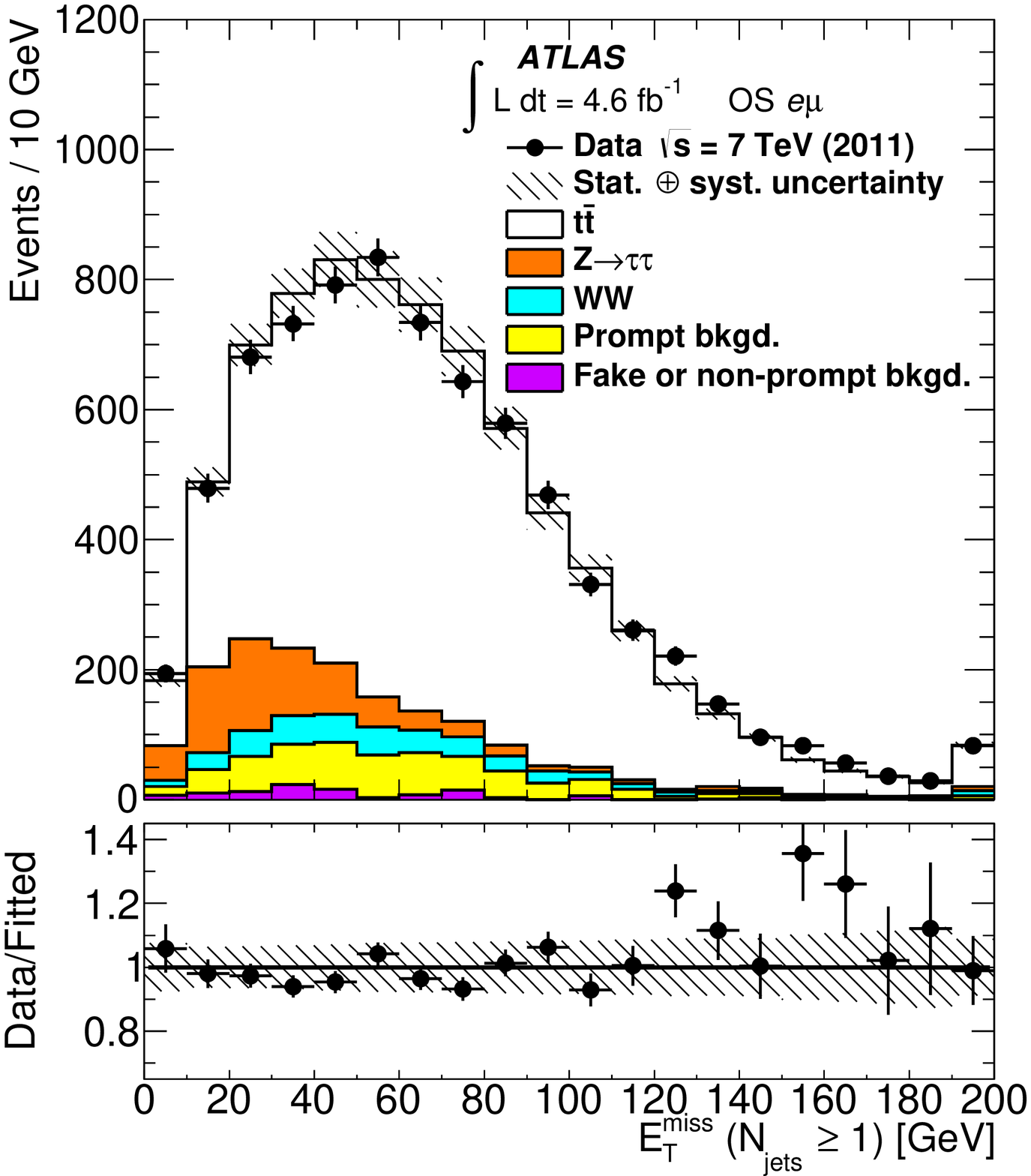} \\
      (c) & (d) \\
    \end{tabular}
  \end{center}
  \caption{Comparison between data and Monte Carlo samples (including the data-driven fake or non-prompt background) after fitting signal processes to data corresponding to an
  integrated luminosity of \lumi{} for events producing one electron and one muon of OS charge: (a) \njets{}, with bins corresponding to 0-jets and $\geq 1$-jet; (b) missing transverse momentum, \met{};
    (c) \met{} for $N_{\rm jets} = 0$  and (d) \met{} for $N_{\rm jets} \geq 1$.
    The electron and muon satisfy the signal region selection criteria presented in Sec.~\ref{Sec:ObjEvtSelec}.
    The hatched regions represent the combination of statistical and systematic uncertainties as described in Table~\ref{t:sys_error} (except for shape uncertainties).
    The last bins in (b), (c) and (d) contain overflow events.
  \label{fig:dist_os_postfit} }
\end{figure*}

\begin{table*}[htbp]
  \caption
      {Expected and observed inclusive yields for events producing one electron and one muon of OS electric charge in an integrated luminosity of \lumi{} at ${\sqrt{s} = 7}$~\TeV.
        The total yields are given followed by the yields subdivided into events producing zero jets and events producing one or more jets with ${\pt > 30}$~\GeV.
        In the final two columns the total yields are subdivided into events that produce \met$ < 30$~\GeV{} and events that produce \met$\geq 30$~\GeV.
        Uncertainties are a quadratic sum of statistical and systematic (including theoretical cross-section) uncertainties, but do not include shape systematic uncertainties.
	The net predicted yields are calculated using unrounded contributions.}
      \label{t:yield_os}
      \begin{ruledtabular}
      \begin{tabular}{lrrrrr}
        Process         &
        \multicolumn{1}{c}{Total}          &
        \multicolumn{1}{c}{$\njets = 0$}   &
        \multicolumn{1}{c}{$\njets \geq 1$}&
        \multicolumn{1}{c}{$\met < 30$~\GeV}&
        \multicolumn{1}{c}{$\met \geq 30$~\GeV}\\ \hline
        & & & & & \\ [-1.5ex]
        $t\bar{t}$      &
        $5900   \pm     500$    &
        $\phantom{0}230$              &
        $5670$          &
        $\phantom{0}860$             &
        $5100   $       \\
        $WW$    &
        $1400\pm 100    $       &
        $1030   $       &
        $\phantom{0}360 $       &
        $\phantom{0}420 $       &
        $\phantom{0}970 $       \\
        $Z\rightarrow\tau\tau$  &
        $3500   \pm     250$    &
        $2900   $       &
        $\phantom{0}610 $     &
        $3000   $       &
        $\phantom{0}520  $     \\
        Single top      &
        $\phantom{0}590 \pm     \phantom{0}50$  &
        $\phantom{00}80 $  &
        $\phantom{0}510 $       &
        $\phantom{00}90 $  &
        $\phantom{0}500 $       \\
        $WZ/ZZ$ &
        $\phantom{00}90 \pm     \phantom{0}40$  &
        $\phantom{00}30 $       &
        $\phantom{00}60 $       &
        $\phantom{00}30 $       &
        $\phantom{00}60 $       \\
        Fake or non-prompt &
        $\phantom{0}210 \pm     170$    &
        $\phantom{0}110 $     &
        $\phantom{0}100 $     &
        $\phantom{00}50 $     &
        $\phantom{0}160 $       \\\hline
        & & & & & \\ [-1.5ex]
        Predicted &
        $11700  \pm     600$    &
        $4400   $       &
        $7300   $       &
        $4400   $       &
        $7300   $       \\ \hline
        & & & & & \\ [-1.5ex]
        Observed &
        \multicolumn{1}{c}{12224}       &
        \multicolumn{1}{r}{4744}     &
        \multicolumn{1}{r}{7480}     &
        \multicolumn{1}{r}{\phantom{000}4750}     &
        \multicolumn{1}{r}{\phantom{000}7474}     \\
	\end{tabular}
\end{ruledtabular}
\end{table*}

\begin{table*}[htbp]
  \caption
      {Fitted and observed inclusive yields for events producing one electron and one muon of OS electric charge in an integrated luminosity of \lumi{} at ${\sqrt{s} = 7}$~\TeV.
        The total yields are given followed by the yields subdivided into events producing zero jets and events producing one or more jets with ${\pt > 30}$~\GeV.
        In the final two columns the total yields are subdivided into events that produce \met$ < 30$~\GeV{} and events that produce \met$\geq 30$~\GeV.
        Uncertainties are a quadratic sum of statistical and systematic uncertainties.
	The net fitted yields are calculated using unrounded contributions.}
      \label{t:yield_os_fitted}
      \begin{ruledtabular}
      \begin{tabular}{lrrrrr}
        Process          &
        \multicolumn{1}{c}{Total}           &
        \multicolumn{1}{c}{$\njets = 0$}    &
        \multicolumn{1}{c}{$\njets \geq 1$} &
        \multicolumn{1}{c}{$\met < 30$~\GeV} &
        \multicolumn{1}{c}{$\met \geq 30$~\GeV} \\ \hline
        & & & & & \\ [-1.5ex]
        $t\bar{t}$    &
        $6050    \pm    350$    &
        $\phantom{0}240$     &
        $5810$    &
        $\phantom{0}880$     &
        $5170$    \\
        $WW$    &
        $1480    \pm     220$ &
        $1120$    &
        $\phantom{0}360$    &
        $\phantom{0}450$    &
        $1030$    \\
        $Z\rightarrow\tau\tau$    &
        $3840    \pm    300$ &
        $3170$&
        $\phantom{0}670$     &
        $3280$    &
        $\phantom{0}560$     \\

        Single top    &
        $\phantom{0}590    \pm    \phantom{0}50$    &
        $\phantom{00}80$  &

        $\phantom{0}510$    &
        $\phantom{00}90$  &
        $\phantom{0}500$    \\

        $WZ/ZZ$    &
        $\phantom{00}90    \pm    \phantom{0}40$    &
        $\phantom{00}30$    &
        $\phantom{00}60$    &
        $\phantom{00}30$    &
        $\phantom{00}60$    \\

        Fake or non-prompt &
        $\phantom{0}210    \pm    170$    &
        $\phantom{0}110$     &
        $\phantom{00}100$     &
        $\phantom{00}50$     &

        $\phantom{0}160$    \\\hline
        & & & & & \\ [-1.5ex]
        Fitted &
        $12260    \pm    540$    &
        $4750$&
        $7510$&

        $4780$&
        $7480$\\ \hline
        & & & & & \\ [-1.5ex]
        Observed &
        \multicolumn{1}{c}{12224}    &
        \multicolumn{1}{r}{\phantom{0000}4744}      &
        \multicolumn{1}{r}{\phantom{0000}7480}      &
        \multicolumn{1}{r}{\phantom{0000}4750}      &
        \multicolumn{1}{r}{\phantom{0000}7474}      \\
        \end{tabular}
      \end{ruledtabular}
\end{table*}
In Table~\ref{t:results}, the fitted yields are shown together with the acceptance correction factors \A{} and \C{} introduced in Sec.~\ref{Sec:FitMeth}, the branching ratios $\mathcal{B}$, and the fiducial and full cross-sections calculated using Eqs.~(\ref{eq:fid_xsection}) and~(\ref{eq:tot_xsection}).
For these branching ratios, the most precise available measurements are used~\cite{Beringer:1900zz}, including the best theoretical prediction of the $W$ leptonic branching ratio, $B(W\to\ell \nu) = 0.1082$ with $0.07\%$ uncertainty.
A fiducial cross-section, for which electrons and muons from tau-lepton decays in \ttbar{} and $WW$ are removed, is also quoted along with a ratio, $R_{\mathcal{C}}$, that translates between the two fiducial region definitions.
This additional fiducial definition is implemented to allow comparisons with predictions for \ttbar{} and $WW$ fiducial cross-sections that do not include tau-lepton decays to electrons and muons.
Such a redefinition of the fiducial region does not alter the product $\A \cdot \C$ nor the relative uncertainties on the fiducial cross-sections.
Also shown are the full uncertainties accompanied by a breakdown of the systematic uncertainty into its three main components (discussed in Sec.~\ref{Sec:FitMeth}, namely those arising from normalization, from shape, and from the fake or non-prompt backgrounds).
For the \ttbar{} and \Ztau{} processes, which have higher production rates, the normalization uncertainty is dominant while the shape uncertainty is dominant for the lower-rate \WW{} process.
This shape uncertainty is not shown in Figs.~\ref{fig:dist_os_prefit} and~\ref{fig:dist_os_postfit}, leading to some underestimate of the error bands at high values of \met{} in Figs.~\ref{fig:dist_os_prefit}(c) and~\ref{fig:dist_os_postfit}(c), where the \WW{} process is dominant.

\begin{table*}[htbp]
\caption{Summary of fitted yields (unrounded), acceptance correction factors, and cross-section measurements.
The acceptance correction factors, $\A \cdot \C$ and $\mathcal{C}$, are extracted from simulated events.
The branching ratios are taken from the best theoretical calculations or experimental measurements~\cite{Beringer:1900zz}.
The fiducial and full cross-sections are calculated using Eqs.~(\ref{eq:fid_xsection}) and~(\ref{eq:tot_xsection}) and accompanied by statistical uncertainties, systematic uncertainties, and uncertainties associated with the luminosity and LHC beam energy.
Also given is a breakdown of the systematic uncertainty including template normalization uncertainties, template shape uncertainties, and uncertainties attributed to the estimation of the fake or non-prompt background.
Fiducial cross-sections for \ttbar{} and $WW$ where leptons from $\tau$ decays are excluded from the definition of the fiducial region are also given along with the ratio, $R_{\mathcal{C}}$, used to translate to the ﬁducial region that includes leptons from $\tau$ decays. The factor 
$R_{\mathcal{C}}$ is defined as the ratio between the acceptance when $\tau$ decays are included in
the definition and when $\tau$ decays are not.}\centering
\label{t:results}
\begin{ruledtabular}
\begin{tabular}{lccc}
Process                                   & \ttbar              & $WW$             & \Ztau              \\
&&&\\
Fitted Yield $N_{\mathrm{fit}}$  & $6049$  & $1479$  & $3844$   \\
$\mathcal{C}$          & $0.482$ & $0.505$ & $0.496$ \\
$R_{\mathcal{C}}$    & $1.150$ & $1.133$ &         \\
$\A \cdot \C$         & $0.224$ & $0.187 $ & $0.0115$ \\
Branching Ratio $B$ &  0.0324 &  0.0324  & 0.0621 \\
&&&\\
&&&\\ \hline
&&&\\ 
$\sigma^{\mathrm{fid}}_X$  [fb] & \ttbarXsecFid       & \WWXsecFid       & \ZtauXsecFid	     \\
Statistical                               & \ttbarUncertStatFid & \WWUncertStatFid & \ZtauUncertStatFid \\
Systematic 	                          & $\pm 140$        & $+88(-95)$       &  $+89(-116)$       \\ 
Luminosity                                & \ttbarUncertLumFid & \WWUncertLumFid  &  \ZtauUncertLumFid        \\
LHC beam energy                           & \ttbarUncertBEFid  & \WWUncertBEFid   &  \ZtauUncertBEFid        \\
&&&\\ 
$\sigma^{\mathrm{fid}}_X$ (excluding $\tau\to \ell\nu\nu$)  [fb] & \ttbarXsecFidnotau       & \WWXsecFidnotau       & \\
Statistical                               & \ttbarUncertStatFidnotau & \WWUncertStatFidnotau & \\
Systematic 	                          &  $\pm 120$            & $+78(-84)$       & \\ 
Luminosity                                & \ttbarUncertLumFidnotau & \WWUncertLumFidnotau & \\
LHC beam energy                           & \ttbarUncertBEFidnotau & \WWUncertBEFidnotau & \\
&&&\\ 
Uncertainties (\%) &&&			  \\ 
Statistical      & $1.5$ 	 & $5.0$ 	  & $2.0$ 	 \\ 	
Systematic       & $5.1$         & $+13.7(-14.9)$ & $+5.5(-7.0)$ \\
Luminosity       & $1.8$         & $1.8$          & $1.8$        \\
LHC beam energy  & $1.8$         & $1.0$          & $0.8$        \\
Total            & $5.9$         & $15.9$         & $7.5$        \\
&&&\\ 
Breakdown of systematic uncertainty (\%) &&& \\	
Normalization  & $+4.6(-4.3)$  & $4.3(-3.8)$    & $+4.2(-3.9)$ \\	        
Shape          & $+1.8(-2.4)$  & $+11.7(-13.2)$ & $+3.0(-5.6)$ \\
Fake or non-prompt background & $\pm 0.8$     & $\pm 5.6$      & $\pm 0.7$    \\
&&&\\ 
&&&\\ \hline
&&&\\ 
$\sigma^{\mathrm{tot}}_X$ [pb]              & \ttbarXsecTot       & \WWXsecTot       & \ZtauXsecTot    \\
Statistical                               & \ttbarUncertStat    & \WWUncertStat    & \ZtauUncertStat \\
Systematic 	                          & $+9.7(-9.5)$        & $+7.3(-8.0)$     & $+72(-88)$        \\ 
Luminosity                                & \ttbarUncertLum     & \WWUncertLum     & \ZtauUncertLum       \\
LHC beam energy                           & \ttbarUncertBE      & \WWUncertBE      & \ZtauUncertBE        \\
&&&\\ 
Uncertainties (\%) &&&			  \\ 
Statistical & $1.5$ 	    & $5.0$ 	     & $2.1$ 	    \\ 	
Systematic  & $+5.4(-5.3)$  & $+13.8(-14.9)$ & $+6.1(-7.5)$ \\
Luminosity       & $1.8$         & $1.8$          & $1.8$        \\
LHC beam energy  & $1.8$         & $1.0$          & $0.8$        \\
Total            & $6.1$         & $15.9$         & $8.0$        \\
&&&\\ 
Subdivision of systematic uncertainty (\%) &&& \\	
Normalization  & $+4.7(-4.3)$  & $+4.2(-3.7)$   & $+5.1(-4.6)$ \\	        
Shape          & $+1.8(-2.4)$  & $+11.7(-13.2)$ & $+3.0(-5.6)$ \\
Fake or non-prompt background & $\pm 0.8$     & $\pm 5.6$      & $\pm 0.7$    \\
\end{tabular}
\end{ruledtabular}
\end{table*}

\subsection{Comparison to previous ATLAS measurements}

This analysis is the first simultaneous measurement of the \ttbar{}, \WW{}, and \Ztau{} cross-sections at $\sqrt{s}=7$~\TeV{}.
Measured cross-sections are summarized and compared to previous measurements and predictions in \mbox{Table~\ref{tab:results_comparison}}.
The \ttbar{} cross-section obtained from the simultaneous measurement is in agreement with
the dedicated \ttbar{} cross-section measurement in the dilepton channel~\cite{Aad:2014kva} at $\sqrt{s}=7$~\TeV{} with identical
integrated luminosity. The dedicated measurement benefits from a more optimised electron identification which reduces
the overall systematic uncertainty associated with the measurement. 
Both measurements assume a top quark mass of $172.5 \GeV$; in the simultaneous measurement the dependence of the measured cross-section on the assumed mass is found to be $-0.8~\mathrm{pb}\, /\GeV$.

In the \WW{} channel, the dedicated analysis at $\sqrt{s}=7$~\TeV{}~\cite{CERN-PH-EP-2012-242} with an integrated luminosity of \lumi{} has significantly greater precision as a result of large shape uncertainties in the simultaneous measurement.
As the smallest of the three measured signals, \WW{} is the one subject to the largest relative variations in the simultaneous fit and has large uncertainties.

Finally, the \Ztau{} simultaneous measurement shows smaller uncertainties than the dedicated measurement~\cite{Aad:2011kt} at $\sqrt{s}=7$~\TeV{} with an integrated luminosity of 36~\ipb{}.
Statistical and luminosity uncertainties are substantially smaller due to the larger data sample with a more precise luminosity determination.

The measurements presented here include the effect of the uncertainty on the LHC beam collision energy, which was not evaluated in prior measurements.
Overall, the comparisons show that each simultaneous cross-section measurement is consistent with its corresponding dedicated ATLAS measurement.

\subsection{Comparison to theoretical calculations}

Figures~\ref{f:fit_errorcontours_fiducial} and~\ref{f:fit_errorcontours} show the best-fit cross-section values with likelihood contours obtained from the simultaneous fit, overlayed with theoretical cross-section predictions. These do not include the contribution from leptonically decaying taus.
The numerical correlation values from the likelihood 
fit are given in Table~\ref{tab:correlation_factors} for each pair of signal processes.
These values give the correlations between the numbers of fitted events in the fiducial region.

NLO fiducial and NLO full cross-section predictions were computed using \mcfm{} v6.6~\cite{MCFM} except 
for the \Ztau{} fiducial cross-section, which was computed with \mcnlo{} interfaced to \herwig{}, \tauola{} and \photos{}.
The computed \WW{} cross-section does not include the contribution from $gg\to H \to\WW$, which is expected to contribute roughly 5\% of the total \WW{} cross-section as discussed in Sec.~\ref{Sec:MC} B.
Fiducial calculations are performed for the region excluding electrons or muons from tau-lepton decays. 

Theoretical predictions were calculated
for the following PDF sets: ABM11~\cite{Alekhin:2012ig}, MSTW2008CPdeut~\cite{Martin:2012da}, CT10, HERAPDF15~\cite{Radescu:2013mka}, NNPDF2.3~\cite{Ball:2012cx}, JR09~\cite{JimenezDelgado:2009tv} (for NNLO calculations) and epWZ~\cite{Aad:2012sb} (for NNLO calculations). In both figures, the markers represent the cross-sections
calculated for a pair of processes using a specific central PDF with its error bars depicting the uncertainty due to
the choice of renormalization ($\mu_{\mathrm{R}}$) and factorization ($\mu_{\mathrm{F}}$) scales.
No attempt is made to treat these scale choices in a correlated way between processes.
The asymmetric scale uncertainty is obtained from the maximum upper and lower deviation from the central value ($\mu_{\mathrm{R}}$ and $\mu_{\mathrm{F}}$) found in a process-specific grid composed of seven cross-sections.
These were calculated by independently varying values of $\mu_{\mathrm{R}}$ and $\mu_{\mathrm{F}}$ by factors of 1/2, 1 and 2 (while ignoring the
cases where $\mu_{\mathrm{R}}$ is doubled and $\mu_{\mathrm{F}}$ is halved and vice versa). 
The central values of $\mu_{\mathrm{R}}$ and $\mu_{\mathrm{F}}$ are set to process-specific values: $m_{t}$ for \ttbar{}, $m_{W}$ for \WW{}, and $m_{Z}$ for \Ztau{}.

The theory contours shown in Fig.~\ref{f:fit_errorcontours_fiducial} correspond to the 68\% confidence level (CL) regions around
each cross-section prediction calculated from
the error sets associated with each specific PDF (intra-PDF uncertainties, defined in Sec.~\ref{Sec:FitMeth}).
The derived uncertainties from different PDF sets are scaled so that all the contours reflect a 68\% CL 
and are constructed using prescribed recipes (in the case of the HERAPDF15 the contour displays asymmetrical errors). 

The fiducial cross-sections provide the most direct comparison between theory and experiment.
Since the fiducial region is chosen to correspond to the sensitive volume of the detector, the theoretical uncertainties are small on the measured values of the fiducial cross-sections.
The uncertainty regions in the fiducial measurements in Fig.~\ref{f:fit_errorcontours_fiducial} suggest that the NLO predictions underestimate all three cross-sections, especially in the case of \Ztau{} versus \ttbar{}, irrespective of the PDF model.
The \WW{} fiducial measurement, however, is consistent with predictions from each PDF model considered, especially considering the fact that the theory predictions in Fig.~\ref{f:fit_errorcontours_fiducial} do not account for the $gg\to H\to\WW{}$ contribution and therefore underestimate the fiducial cross-section by approximately 5\% (see Sec.~\ref{Sec:MC} C).

Full cross-section measurements are shown in Fig.~\ref{f:fit_errorcontours} accompanied by 68\% CL and 90\% CL contours calculated for the case where the fit only includes the
theoretical uncertainty (inner contours) and the case when the full uncertainty is included (outer contours).
Although larger acceptance uncertainties clearly reduce the separation power with respect to the fiducial measurements, here the full theoretical calculations at NNLO in QCD can be used for \Ztau{} versus \ttbar{}, as shown in Fig.~\ref{f:fit_errorcontours}(d).
As described in Sec.~\ref{Sec:MC}, the software packages {\sc FEWZ} and \toppp{} were used to calculate the cross-sections to NNLO.
Figure~\ref{f:fit_errorcontours}(d) (NNLO case) in contrast to Fig.~\ref{f:fit_errorcontours}(c) (NLO case) shows good overlap between the experimental measurement and most of the NNLO theoretical predictions and corresponding PDF sets for \Ztau{} versus \ttbar{} where they are available.
Also notable is the difference in the uncertainties in theoretical predictions: in the NLO case scale uncertainties are dominant, while in the NNLO case the PDF model provides the dominant uncertainty.
Theory contours using ABM11 and JR09 PDFs, however, do not overlap with the measurements.
For the former, one significant reason for a lower \ttbar{} cross-section 
lies in the value of $\alpha_s$ employed in its calculation.
At NNLO its value is 0.113, which is substantially
lower than the range of 0.117 to 0.118 employed by most of the other PDF models here.
In the case of JR09, which is only considered in the comparison of NNLO calculations, the 5\% difference in the \Ztau{} cross-section is consistent with what is reported elsewhere~\cite{JimenezDelgado:2009tv}.

\begin{table*}
\centering
\caption[Cross-sections compared with theory and dedicated measurements]
{Comparisons of the total \ttbar{}, \WW{}, and \Ztau{} cross-sections as measured simultaneously in this analysis with symmetrized uncertainties to previous dedicated ATLAS measurements and to the most accurate predictions from QCD.
The NLO QCD prediction for \WW{} presented here is the sum of the $qq\to\WW$, $gg\to\WW$, and $gg\to H\to\WW$ cross-sections.
The ATLAS dedicated \Ztau{} production cross-section was measured in the fiducial region where $66$~\GeV$ < m_{\tau\tau} < 116$~\GeV{} and so is corrected by a factor $1.1$ to compare it directly with 
the \Ztau{} cross-section measured here in the fiducial region $m_{\tau\tau} > 40$~\GeV.
}
\label{tab:results_comparison}
\begin{ruledtabular}
\renewcommand{\arraystretch}{1.2}
\begin{tabular}{llcccccccc}
Process                    & Source & $\sigma^{\mathrm{tot}}_X$ & \multicolumn{5}{c}{Uncertainties}               &  $\int \mathcal{L}\,\mathrm{d}t$ & Reference  \\ 
                           &        & [pb]                         & Stat.  & Syst.  & Lumi.  & Beam  & Total  &  [$\mathrm{fb}^{-1}$] & \\ \hline
\multirow{3}{*}{\ttbar}   & Simultaneous  & 181  & 3  & 10 & 3 & 3 & 11 &  4.6                  & \\ 
                          & Dedicated     & 183  & 3  &  4 & 4 & 3 &  7 &  4.6  &  \cite{Aad:2014kva} \\ 
                          & NNLO QCD & 177  &      &     &     &     & 11    &        &  \cite{Czakon:2013goa} \\  \hline   
\multirow{3}{*}{$WW$}     & Simultaneous                                             & 53.3 & 2.7  & 7.7 & 1.0 & 0.5 & 8.5  & 4.6  & \\
                          & Dedicated                                                & 51.9 & 2.0  & 3.9 & 2.0 &     &  4.9 & 4.6  & \cite{CERN-PH-EP-2012-242}\\ 
                          & NLO QCD & 49.2 & &    &     &     &  2.3    &      &  \cite{Campbell:2009kg}\\ \hline
\multirow{3}{*}{\Ztau{}} & Simultaneous                                             & 1174  &  24  &  80  & 21  &    9  & 87  & 4.6 &   \\ 
                          & Dedicated ($e\mu$)                                       & 1170  & 150  &  90  & 40  &      & 170  & 0.036 &   \cite{Aad:2011kt} \\ 
                          & NNLO QCD & 1070  &    &   &     &      & 54  &     &   \cite{FEWZ2011} \cite{MSTW2008}\\
\end{tabular}
\end{ruledtabular}
\end{table*}

\begin{figure*}[htbp]
  \centering
  \begin{tabular}{cc}
    \includegraphics[width=0.48\textwidth]{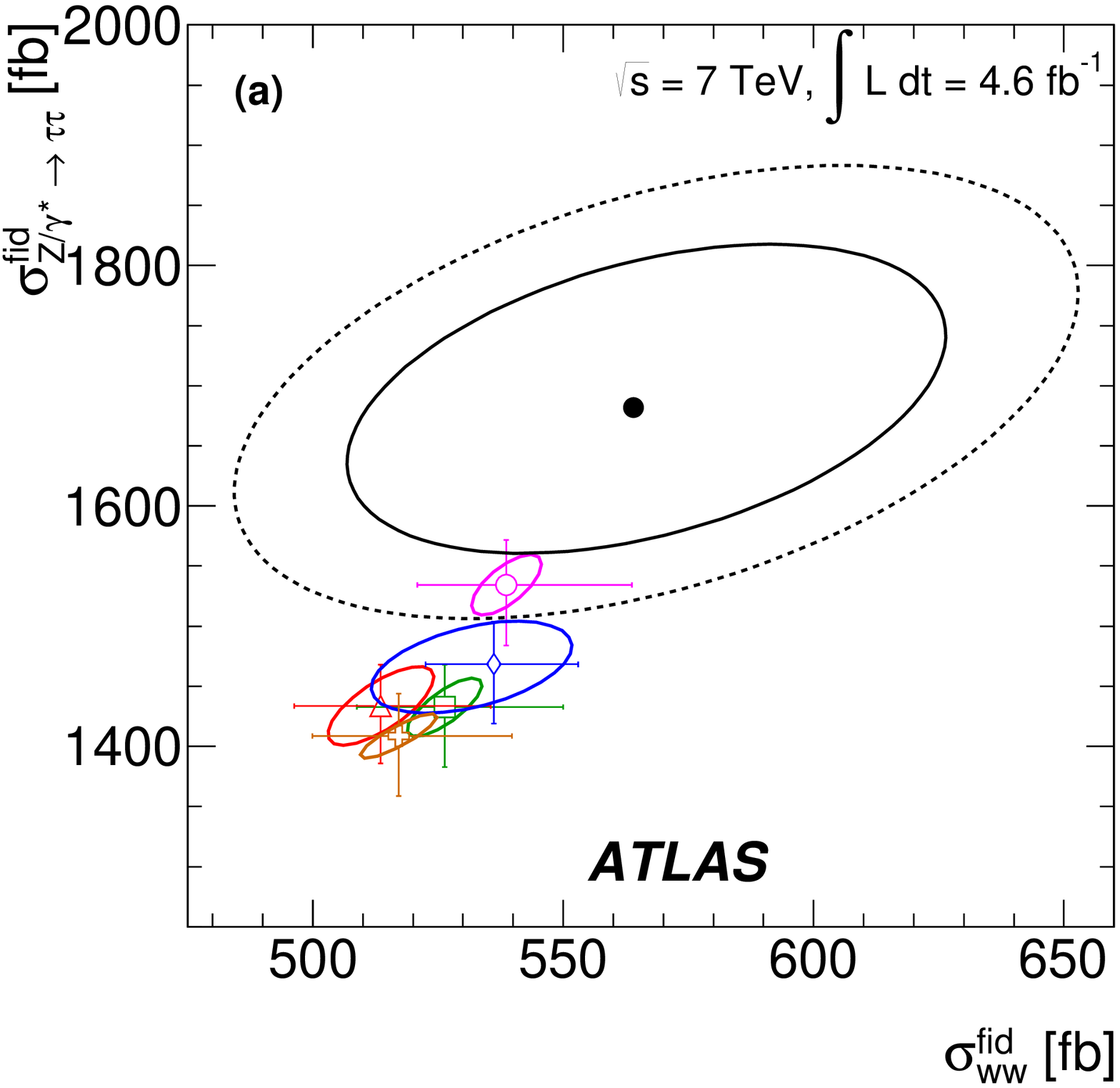} &  
    \includegraphics[width=0.48\textwidth]{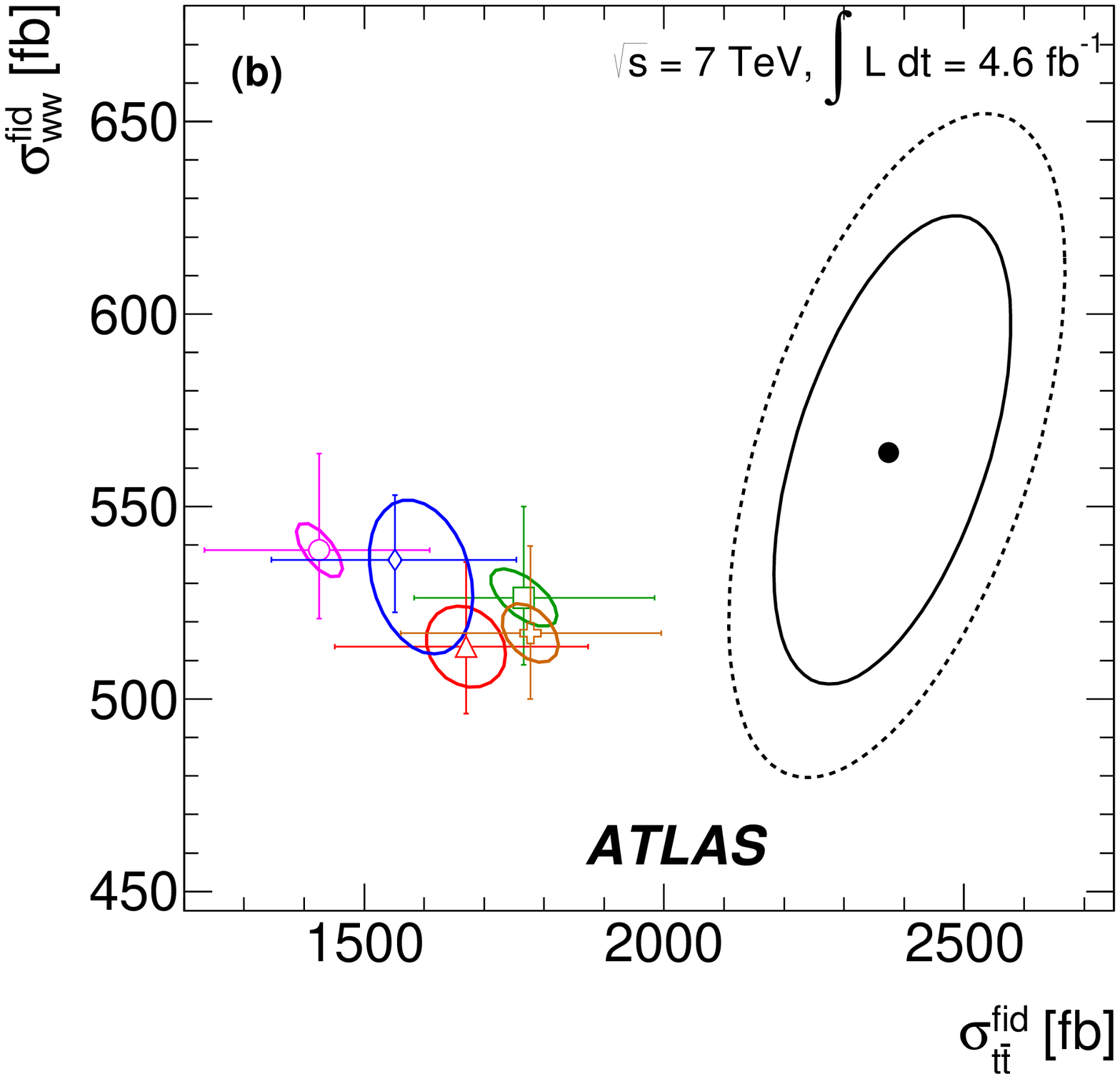} \\
    (a) & (b) \\
    \includegraphics[width=0.48\textwidth]{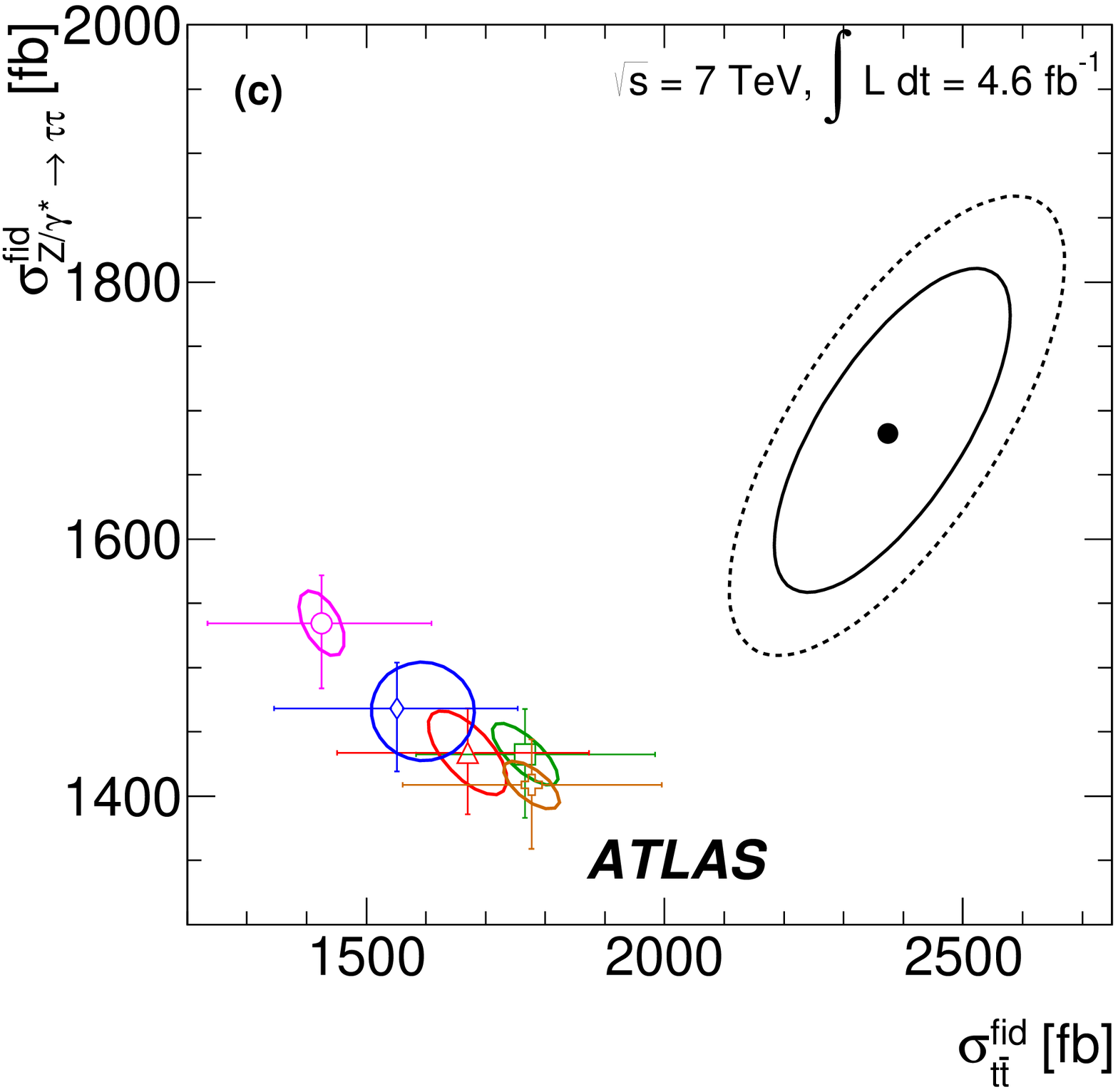} &
    \includegraphics[width=0.48\textwidth]{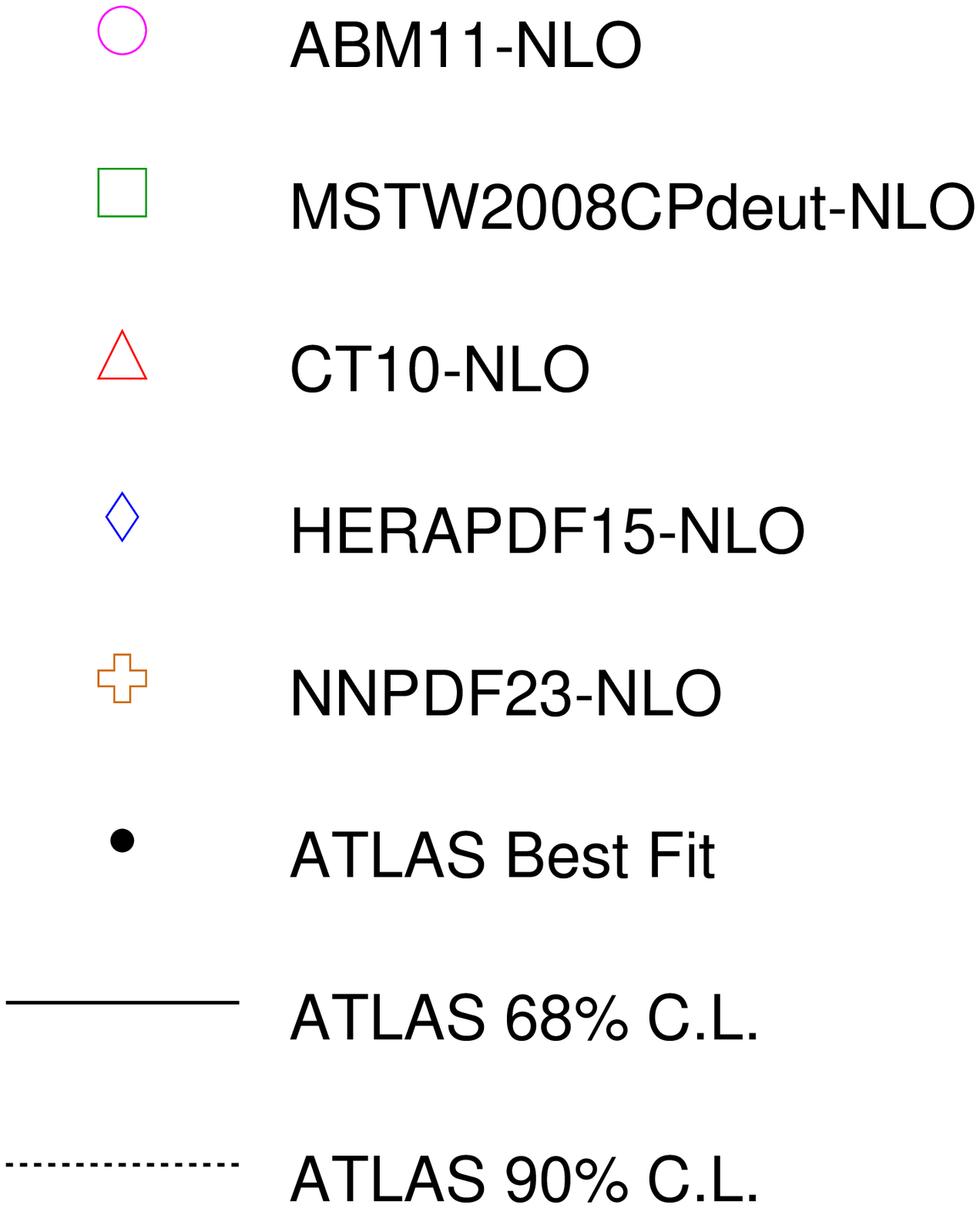} \\
    (c) & \\
  \end{tabular}
  \caption{Contours of the likelihood function as a function of two fiducial production cross-sections of interest:
    (a) $\sigma^{\mathrm{fid}}_{\Ztau}$ versus $\sigma^{\mathrm{fid}}_{\WW}$,
    (b) $\sigma^{\mathrm{fid}}_{\WW}$ versus $\sigma^{\mathrm{fid}}_{\ttbar}$,
    (c) $\sigma^{\mathrm{fid}}_{\Ztau}$ versus  $\sigma^{\mathrm{fid}}_{\ttbar}$ .
The contours obtained from the data (full circle) represent the 68\% CL (full line) and 90\% CL (dashed line) areas accounting for the full set of systematic uncertainties described in Table~\ref{t:sys_error}.
The fiducial cross-sections for \WW{} and \ttbar{} exclude contributions from tau-lepton decays.
The theoretical \WW{} cross-section does not include contributions from $gg\to H \to \WW$.
The theoretical fiducial cross-section predictions are shown at next-to-leading-order (NLO) in QCD for different PDF sets (open symbols) with the ellipse contours corresponding to the 68\% CL uncertainties on each PDF set.
Also shown as horizontal and vertical error bars around each prediction are the uncertainties due to the choice of QCD factorization and renormalization scales (see text).
}
  \label{f:fit_errorcontours_fiducial} 
\end{figure*}

\begin{figure*}[htbp]
  \centering
  \begin{tabular}{cc}
    \includegraphics[width=0.48\textwidth]{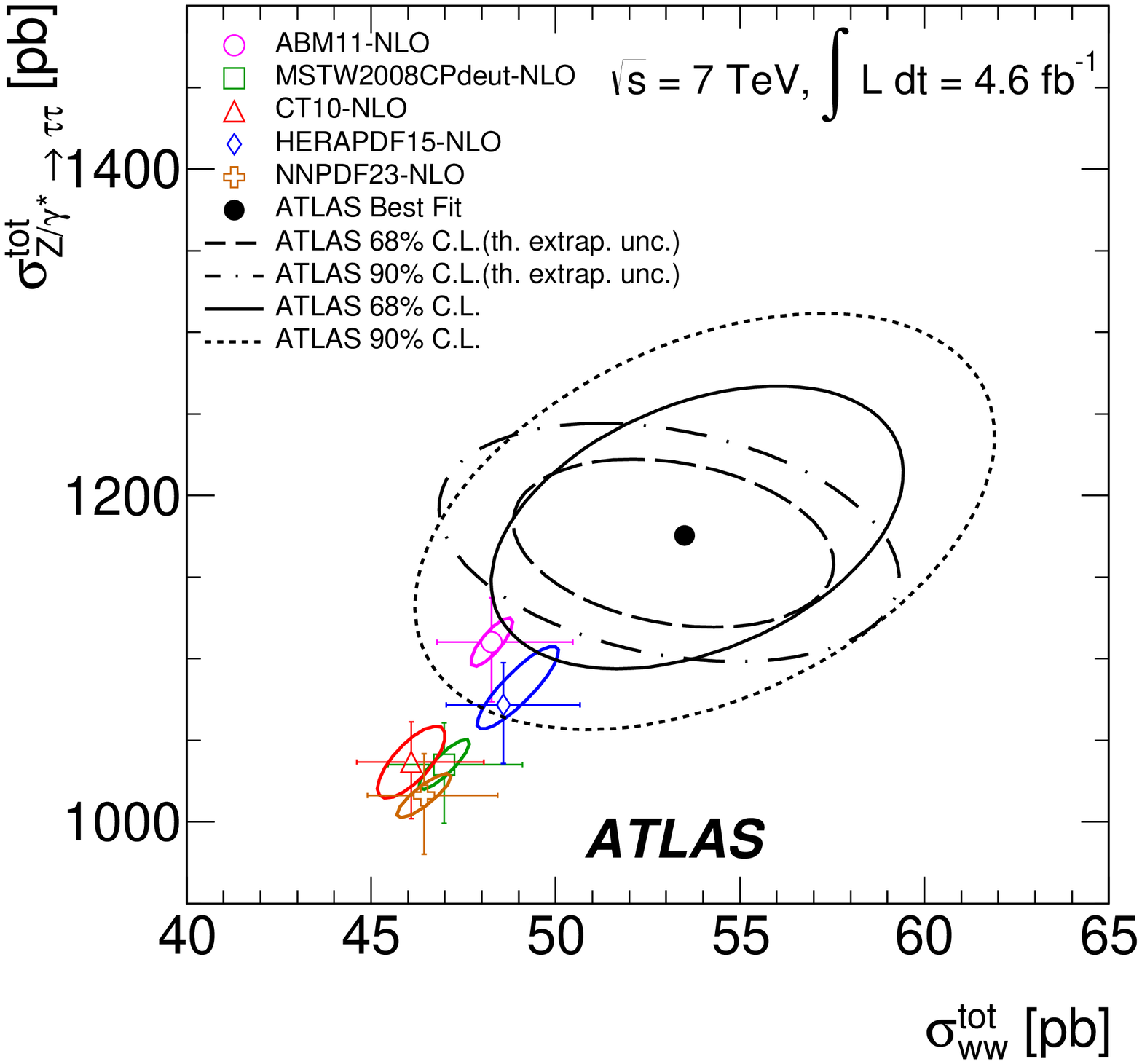} &
    \includegraphics[width=0.48\textwidth]{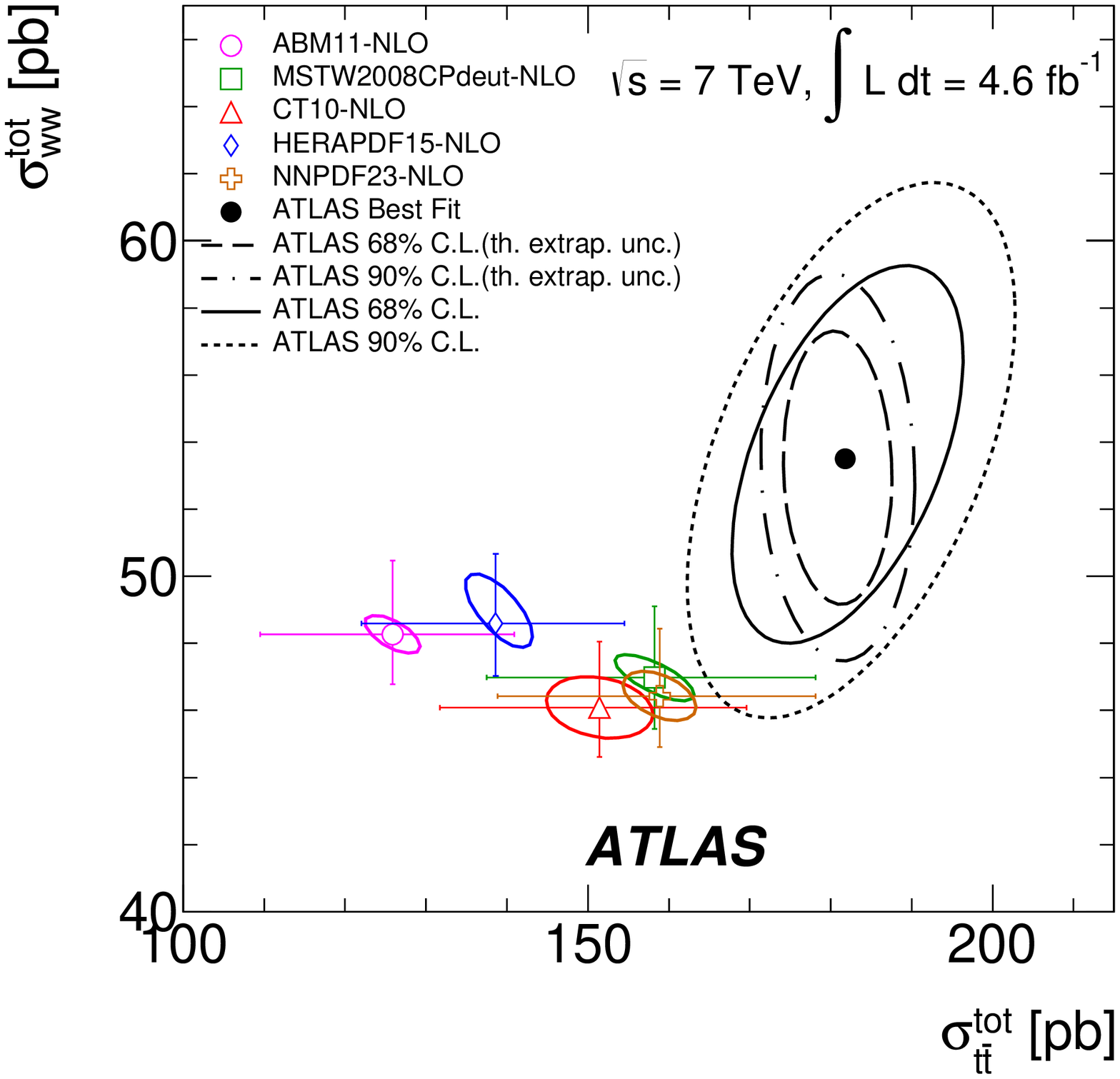} \\
    (a) & (b) \\
    \includegraphics[width=0.48\textwidth]{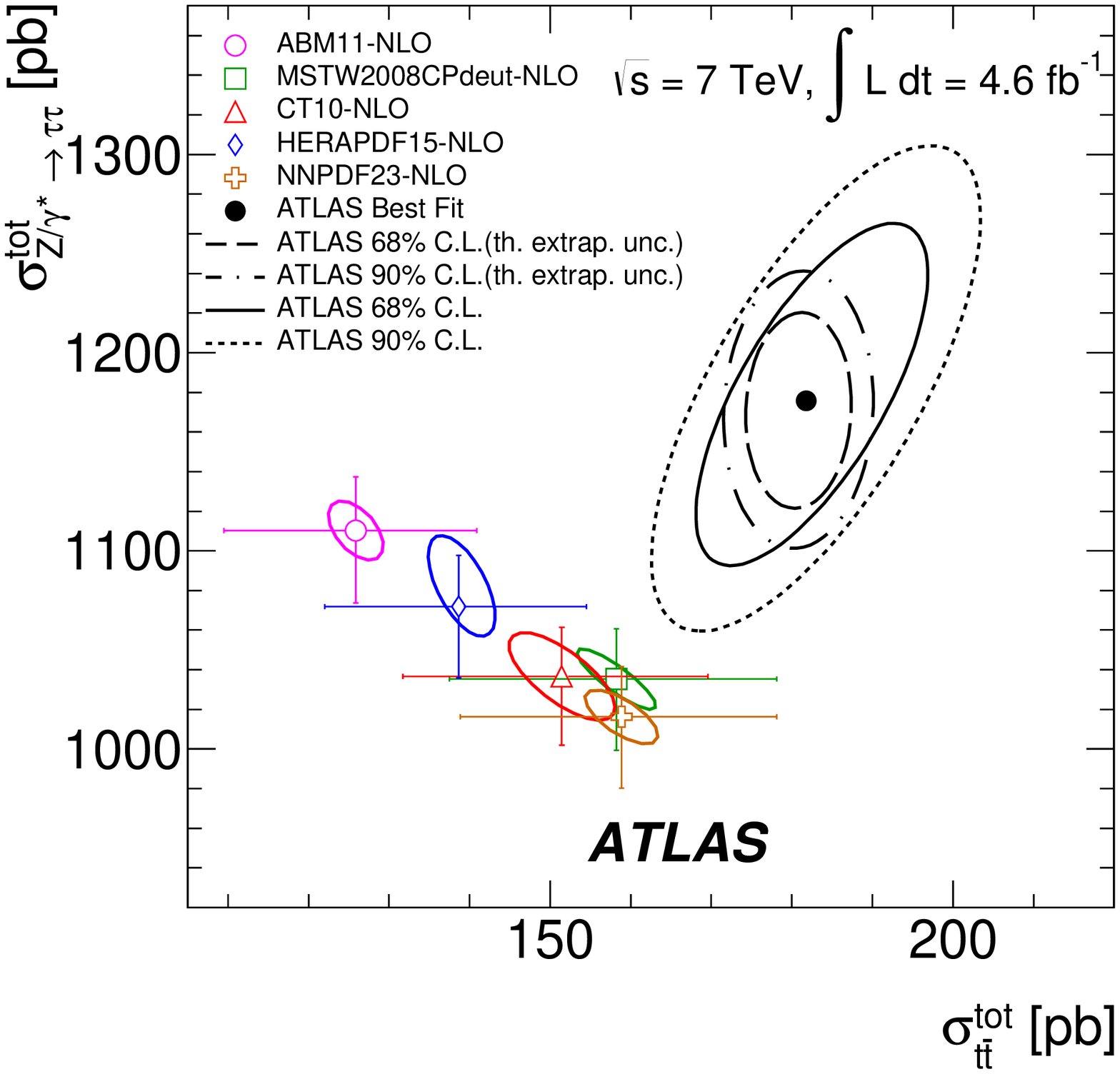} &
    \includegraphics[width=0.48\textwidth]{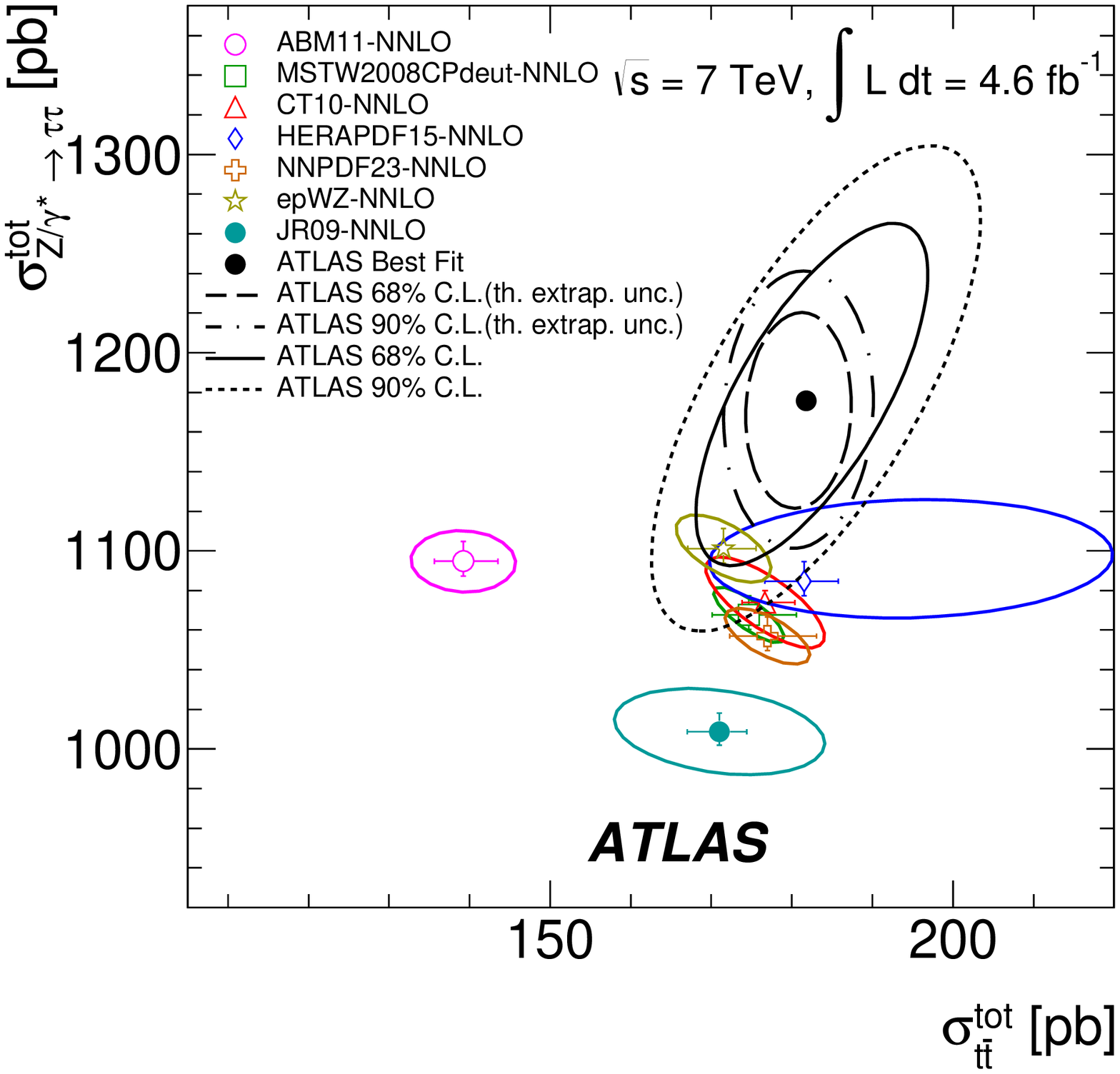} \\
    (c) & (d) \\
  \end{tabular}
  \caption{Contours of the likelihood function as a function of two full production cross-sections of interest:
    (a) $\sigma^{\mathrm{tot}}_{\Ztau}$ versus $\sigma^{\mathrm{tot}}_{WW}$ compared to NLO predictions;
    (b) $\sigma^{\mathrm{tot}}_{WW}$ versus $\sigma^{\mathrm{tot}}_{\ttbar}$ compared to NLO predictions;
    (c,d)  $\sigma^{\mathrm{tot}}_{\Ztau}$ versus $\sigma^{\mathrm{tot}}_{\ttbar}$ compared to NLO, NNLO predictions.
The contours obtained from the data (full circle) represent the 68\% CL (full line) and 90\% CL (dashed line) areas accounting for the full set of systematic uncertainties described in Table~\ref{t:sys_error}.
Contours labeled ``th. extrap. uncertainty'' depict the theoretical
uncertainties on extrapolating the fiducial cross-section to the full
phase space and are obtained by constructing a likelihood function
with only theoretical uncertainties.
The theoretical \WW{} cross-section does not include contributions from $gg\to H \to \WW$.
The theoretical cross-section predictions are shown at NLO (a, b, and c) or NNLO (d) in QCD for different PDF sets (open symbols) with the ellipse contours corresponding to the 68\% CL uncertainties on each PDF set.
Also shown as horizontal and vertical error bars around each prediction are the uncertainties due to the choice of QCD factorization and renormalization scales (see text).
    \label{f:fit_errorcontours} } 
\end{figure*}

\begin{table}
\centering
\caption[Measured correlation factors]
{
  Correlation factors of the fitted yields for measured signal processes.
  These values give the correlations between the numbers of fitted events in the fiducial region which includes leptons from $\tau$ decays.
}
\label{tab:correlation_factors}
\begin{ruledtabular}
\begin{tabular}{lc}
Processes      & Correlation \\ \hline
      &  \\ [-0.9ex]
\Ztau{} versus \WW{}    & 0.37        \\ 
\WW{} versus \ttbar{}  & 0.53        \\
\Ztau{} versus \ttbar & 0.61        \\
\end{tabular}
\end{ruledtabular}
\end{table}
\section{Conclusion}
\label{Sec:conclusion}
Simultaneous measurements of the \ttbar{}, \WW{} and \Ztau{} fiducial and total production cross-sections using \lumi{} of data collected with the ATLAS detector from $pp$ collisions at ${\sqrt{s}=7\,\TeV}$ at the LHC are presented.
Exactly two high transverse momentum isolated leptons are selected, and are required to be one electron and one muon of opposite charge.
The number of signal events is extracted using a template fit to the distribution of missing transverse momentum and jet multiplicity observed in the data.The measurements are consistent with the previously published dedicated ATLAS cross-section measurements and with the predicted theoretical cross-sections within their uncertainties.
This simultaneous extraction of the cross-sections for these processes at the LHC 
provides a broader test of the SM predictions than individual measurements by unifying the fiducial region, object and event requirements, and background estimations.
The uncertainty bands of the measured fiducial cross-sections of \ttbar{} and \Ztau{} suggest that the NLO predictions underestimate the data, while comparisons to NNLO calculations indicate that MSTW2008, CT10, HERAPDF, NNPDF, and epWZ PDF sets describe the data well.

\section*{Acknowledgments}

We thank CERN for the very successful operation of the LHC, as well as the
support staff from our institutions without whom ATLAS could not be
operated efficiently.

We acknowledge the support of ANPCyT, Argentina; YerPhI, Armenia; ARC,
Australia; BMWF and FWF, Austria; ANAS, Azerbaijan; SSTC, Belarus; CNPq and FAPESP,
Brazil; NSERC, NRC and CFI, Canada; CERN; CONICYT, Chile; CAS, MOST and NSFC,
China; COLCIENCIAS, Colombia; MSMT CR, MPO CR and VSC CR, Czech Republic;
DNRF, DNSRC and Lundbeck Foundation, Denmark; EPLANET, ERC and NSRF, European Union;
IN2P3-CNRS, CEA-DSM/IRFU, France; GNSF, Georgia; BMBF, DFG, HGF, MPG and AvH
Foundation, Germany; GSRT and NSRF, Greece; ISF, MINERVA, GIF, I-CORE and Benoziyo Center,
Israel; INFN, Italy; MEXT and JSPS, Japan; CNRST, Morocco; FOM and NWO,
Netherlands; BRF and RCN, Norway; MNiSW and NCN, Poland; GRICES and FCT, Portugal; MNE/IFA, Romania; MES of Russia and ROSATOM, Russian Federation; JINR; MSTD,
Serbia; MSSR, Slovakia; ARRS and MIZ\v{S}, Slovenia; DST/NRF, South Africa;
MINECO, Spain; SRC and Wallenberg Foundation, Sweden; SER, SNSF and Cantons of
Bern and Geneva, Switzerland; NSC, Taiwan; TAEK, Turkey; STFC, the Royal
Society and Leverhulme Trust, United Kingdom; DOE and NSF, United States of
America.

The crucial computing support from all WLCG partners is acknowledged
gratefully, in particular from CERN and the ATLAS Tier-1 facilities at
TRIUMF (Canada), NDGF (Denmark, Norway, Sweden), CC-IN2P3 (France),
KIT/GridKA (Germany), INFN-CNAF (Italy), NL-T1 (Netherlands), PIC (Spain),
ASGC (Taiwan), RAL (UK) and BNL (USA) and in the Tier-2 facilities
worldwide.

\clearpage

\providecommand{\href}[2]{#2}\begingroup\raggedright\endgroup

\onecolumngrid
\clearpage
\begin{flushleft}
G.~Aad$^{\rm 84}$,
B.~Abbott$^{\rm 112}$,
J.~Abdallah$^{\rm 152}$,
S.~Abdel~Khalek$^{\rm 116}$,
O.~Abdinov$^{\rm 11}$,
R.~Aben$^{\rm 106}$,
B.~Abi$^{\rm 113}$,
M.~Abolins$^{\rm 89}$,
O.S.~AbouZeid$^{\rm 159}$,
H.~Abramowicz$^{\rm 154}$
% ...
\bigskip
\\
% ...
$^{11}$ Institute of Physics, Azerbaijan Academy of Sciences, Baku, Azerbaijan\\
% ...
$^{84}$ CPPM, Aix-Marseille Universit{\'e} and CNRS/IN2P3, Marseille, France\\
% ...
$^{89}$ Department of Physics and Astronomy, Michigan State University, East Lansing MI, United States of America\\
% ...
$^{106}$ Nikhef National Institute for Subatomic Physics and University of Amsterdam, Amsterdam, Netherlands\\
% ...
$^{112}$ Homer L. Dodge Department of Physics and Astronomy, University of Oklahoma, Norman OK, United States of America\\
$^{113}$ Department of Physics, Oklahoma State University, Stillwater OK, United States of America\\
% ...
$^{116}$ LAL, Universit{\'e} Paris-Sud and CNRS/IN2P3, Orsay, France\\
% ...
$^{152}$ Institute of Physics, Academia Sinica, Taipei, Taiwan\\
% ...
$^{154}$ Raymond and Beverly Sackler School of Physics and Astronomy, Tel Aviv University, Tel Aviv, Israel\\
% ...
$^{159}$ Department of Physics, University of Toronto, Toronto ON, Canada\\
% ...
\end{flushleft}

\end{document}